\documentclass[useAMS,usenatbib]{mn2e}
\usepackage{txfonts}
\usepackage{graphicx}
\usepackage{pdflscape}
\def\NH2{N$_{\mathrm{H}_{2}}$}
\def\mum{$\mu$m}
\def\grad{^{\circ}}
\def\sun{$_{\odot}$}
\def\Hyp{\textit{Hyper}}
\def\Her{\textit{Herschel}}
\def\thres{$\sigma_{t}$}
\def\bgm{M$_{bg}$}
\def\rms{\textit{r.m.s.}}
\def\Msol{M$_\odot$}
\def\Lsol{L$_\odot$}
\def\Lsun{L$_\odot$}

\title{The initial conditions of stellar protocluster formation. II.  A catalogue of starless and protostellar clumps embedded in IRDCs in the Galactic longitude range $15\grad\leq l\leq55\grad$} 

\author[A. Traficante, G. A. Fuller et al.]{A. Traficante$^{1}$\thanks{e-mail:alessio.traficante@manchester.ac.uk}, G.A. Fuller$^{1}$, N. Peretto$^{2}$, J.E. Pineda$^{3}$, and S. Molinari$^{4}$ \\
$^{1}$Jodrell Bank Centre for Astrophysics, School of Physics and Astronomy, University of Manchester, Oxford Road, Manchester M13 9PL, UK\\
$^{2}$School of Physics and Astronomy, Cardiff University, Queens Buildings, The Parade, Cardiff CF24 3AA, UK\\
$^{3}$Max-Planck-Institut für extraterrestrische Physik (MPE), Germany\\
$^{4}$IAPS - INAF, via Fosso del Cavaliere, 100, I-00133 Roma, Italy}

\begin{document}

\date{}

\pagerange{\pageref{firstpage}--\pageref{lastpage}} \pubyear{2011}

\maketitle

\label{firstpage}

\begin{abstract}
 We present a catalogue of starless and protostellar clumps associated with
 infrared dark clouds (IRDCs) in a 40$\grad$ wide region of the inner
 Galactic Plane ($\vert b\vert\leq1\grad$). We have extracted the far-infrared (FIR) counterparts
 of 3493 IRDCs with known distance in the Galactic longitude range
 $15\grad\leq l\leq55\grad$ and searched for the young clumps using Hi-GAL,
 the survey of the Galactic Plane carried out with the \Her\ satellite.  Each
 clump is identified as a compact source detected at 160, 250 and 350\,\mum.
 The clumps have been classified as protostellar or starless, based on their
 emission (or lack of emission) at 70\,\mum. We identify 1723 clumps, 1056
 (61\%) of which are protostellar and 667 (39\%) starless. These clumps are
 found within 764 different IRDCs, 375 (49\%) of which are only associated
 with protostellar clumps, 178 (23\%) only with starless clumps, and 211
 (28\%) with both categories of clumps.  The clumps have a median mass of
 $\sim$250\Msol\ and range up to $>10^4$\Msol\ in mass and up to
 $10^5$\,\Lsol\ in luminosity.
  
The mass-radius distribution shows that almost 30\% of the starless clumps
identified in this survey could form high-mass stars, however these massive
clumps are confined in only $\simeq4\%$ of the IRDCs.  Assuming a minimum mass
surface density threshold for the formation of high-mass stars, the comparison
of the numbers of massive starless clumps and those already containing
embedded sources suggests an upper limit lifetime for the starless phase of
$\sim10^5$ years for clumps with a mass M$>500$\,\Msol.

\end{abstract}

\section{Introduction}
The star formation process begins in massive clouds with a reservoir of gas and dust sufficient to sustain the creation of a cluster of stars \citep[e.g.][]{Lada03}, which in some clouds may eventually form high-mass stellar objects \citep[e.g.,][]{Rathborne06,Peretto09}. Some of these dense, cold clouds which are not yet dominated by star formation absorb the IR emission of the background radiation and therefore can be observed as dark structures in the mid-IR images \citep[infrared dark clouds, IRDCs, e.g.][]{Simon06}.  These relatively undisturbed clouds are favoured places to study the very early stages of star formation \citep{Peretto09,Peretto&Fuller10,Battersby11,Ragan12}.

Within these clouds, the earliest stages are initially characterised by the formation of a starless, dense clump with a size of $\simeq 1$ pc. The cold dust envelope of these clumps emits in the far-IR (FIR)/sub-mm region of the spectrum, but are still IR-quiet at wavelengths $\leq70$ \mum\ \citep{Elia10,Motte10,Giannini12}. As a protostellar core and then protostar eventually forms within the gravitationally bound starless clumps (the \textit{pre-stellar} clumps) it heats the surrounding envelope. The now protostellar clump becomes visible at wavelengths $\leq70$ \mum\ and its bolometric luminosity increases, sustained by the warm inner core(s). As a consequence, the  protostellar luminosity correlates well with the clump emission observed at 70 \mum\ \citep{Dunham08}. 

Large surveys of the Galactic Plane in the FIR-sub-mm region allow us to make a census of these clumps and the young stellar objects (YSOs) within them across the Galaxy. Such surveys provide the significant samples of sources required to understand the star formation mechanisms across a wide range of mass and luminosity regimes as well as identify rare and/or short-lived classes of objects. Several surveys have been carried out in the past few years at increasing sensitivity and spatial resolution in order to build a statistically significant sample of young stellar objects in the Galactic Plane. The APEX Telescope Large Area Survey of the Galaxy \citep[ATLASGAL,][]{Schuller09} and the Bolocam Galactic Plane Survey \citep[BGPS,][]{Aguirre10} have mapped the inner Galactic Plane in the sub-mm range, at 870 \mum\ and 1.1 mm respectively, allowing a census of the sub-mm thermal emission from high-mass regions.  
However, unprecedented opportunities to make a multi-wavelength FIR/sub mm study of the sky arrived with the \Her\ satellite \citep{Pilbratt10}, opening a new window of our understanding of the cold universe and, in this context, Galactic star formation.
For example, the earliest phases of star formation associated with IRDCs have been studied with the EPoS \Her\ key program, for both low-mass \citep{Launhardt13} and high-mass objects \citep{Ragan12}. This survey studied a sample of 12 low-mass and 45 high-mass regions respectively in order to characterise the temperature and column densities of the cores and clumps at different mass regimes.

A comprehensive \Her\ mapping of the Galactic cold and dusty regions has been recently completed with the \Her\ infrared Galactic Plane survey \citep[Hi-GAL,][]{Molinari10_PASP}, which has mapped the whole Galactic Plane at latitudes $\vert b\vert\leq1\grad$ and following the Galactic warp in the wavelength range $70\leq\lambda\leq500$ \mum. 
 
The aim of this work is to produce a first extensive catalogue of young clumps embedded in IRDCs combining the Hi-GAL data with the most comprehensive catalogue of IRDCs to date \citep[][hereafter Paper I]{Peretto09}. This initial catalogue looks in a specific region of the inner Galactic Plane, in the longitude range $15\grad\leq l\leq 55\grad$, which encompasses $\simeq3500$ IRDCs with known distance.

The paper is structured as: in Section \ref{sec:dataset} describes the IRDCs dataset extracted from Paper I and the Hi-GAL data used to extract the FIR/sub-mm IRDCs counterparts; the source extraction and the catalogues generation are described in Section \ref{sec:source_ext_phot}. In this Section we also briefly describe \Hyp, a new algorithm used in this work and developed for source extraction and photometry in complex backgrounds and crowded fields \citep{Traficante14_Hyp}. In Section \ref{sec:mass_luminosity} we describe the statistical distributions of the starless and protostellar clumps properties. Finally, in Section \ref{conclusion} we draw up our conclusions about this first release of clumps in IRDCs.

%
\section{IRDC dataset}\label{sec:dataset}
The IRDCs survey produced in Paper I contains $\simeq11000$ IRDCs identified in absorption against the warm background in the GLIMPSE 8 \mum\ survey of the Galactic Plane \citep{Benjamin03} delimited by $\vert l\vert\leq 65^{\circ}$, $\vert b\vert\leq 1^{\circ}$. The IRDCs have been identified as connected regions with column density higher than \NH2 $\geq1\times10^{22}$ cm$^{-2}$ and a diameter greater than 4\arcsec\ (Paper I). 

We focused on a subsample of 3659 IRDCs observed in the region $15\grad\leq l\leq 55\grad$, $\vert b\vert\leq 1^{\circ}$. This region has been selected for its overlap with the Galactic Ring Survey \citep[GRS,][]{Jackson06}. The GRS emission along the IRDCs line of sight (LOS) has been used to identify clouds and to estimate the IRDCs kinematic distances. The GRS survey mapped the $^{13}$CO J=1-0 emission with the FCRAO 14m telescope with an angular resolution of 46\arcsec\ and a spectral resolution of 0.212 km s$^{-1}$ \citep{Jackson06}. To obtain kinematic distances, we developed an automated procedure that extracts the   $^{13}$CO spectrum towards the centre of each IRDC, identifies the number
of velocity components in it, integrates each component in a 2 km/s interval around the central velocity, and then calculates the ratio map for each of the 
integrated intensity maps with the Herschel column density, smoothed at 46\arcsec\ resolution, centred at the IRDC position. We then select the velocity component for which the ratio map is the flattest, i.e. with a minimum dispersion. In some cases, no $^{13}$CO(1-0) components were found, and a -999 flag was returned for the velocity. We identified 166 clouds with a -999 flag, therefore our starting IRDC catalogue is composed of 3493 clouds. This procedure is simple, and rather easy to implement for such a large number of sources. 
 Kinematical distances were then calculated using the \citet{Reid09} Galactic rotation model, always assuming the clouds located at the near distance in case of a distance ambiguity. 


\subsection{IRDCs Hi-GAL counterparts}\label{sec:HGL_counterparts}
The Hi-GAL survey \citep{Molinari10_PASP} has been carried out using both the PACS \citep[][]{Poglitsch10} and SPIRE \citep[][]{Griffin10} photometry instruments on-board \Her\ \citep{Pilbratt10} in parallel mode, observing the sky at five wavelengths simultaneously (70 and 160 \mum\ with PACS and 250, 350 and 500 \mum\ with SPIRE). The maps have been reduced with the ROMAGAL pipeline \citep{Traficante11}, an enhanced version of the standard \Her\ pipeline specifically designed for Hi-GAL. A weighted post-processing on the maps has been applied to help with image artefact removal \citep{Piazzo11}. The maps have been flux calibrated (by means of an offset subtraction) following the prescription of \citet{Bernard10} and are expressed in MJy/sr. Due to the fast scan-speed and the co-addition on-board \Her\ of 8 samples at 70 \mum\ and 4 samples at 160 \mum, the measured PACS beams in the maps are slightly larger than the nominal ones. The Hi-GAL beams measured on the maps are 10.2$\arcsec$, 13.55$\arcsec$, 18.0$\arcsec$, 24.0$\arcsec$ and 34.5$\arcsec$ at 70, 160, 250, 350 and 500 \mum\ respectively. The map pixel size is 3.2$\arcsec$, 4.5$\arcsec$, 6.0$\arcsec$, 8.0$\arcsec$, 11.5$\arcsec$ at 70, 160, 250, 350 and 500 \mum\ respectively \citep{Traficante11}.

For each IRDCs in our sample we selected the corresponding region in the Hi-GAL 70, 160, 250 and 350 \mum\ maps. We do not include the 500 \mum\ maps in the analysis due to the poor beam resolution compared to the other wavelengths.

Some of the IRDCs of our sample have been already identified in the Hi-GAL survey by \citet{Peretto10}. In this work the authors showed that the IRDC FIR extension is slightly bigger than their appearance at 8 and 24 \mum. Therefore, for each IRDC, we isolate an Hi-GAL region which enlarged the 24 \mum\ IRDC size by the equivalent of a 70 \mum\ beam (10.2$\arcsec$) in each spatial direction. We assume that all the sources extracted between these extended boundaries belong at the same IRDC. Due to the boundaries extension and the proximity of some IRDCs, some sources can be associated with two different clouds. In these cases we associate each source with one IRDC randomly selected from the two clouds.

\section{Clump extraction and photometry}\label{sec:source_ext_phot}
Source extraction and photometry is a well known problem in astronomy, in particular in complex fields such as the Galactic Plane. The high resolution and sensitivity of the new IR instruments have shown the complexity of the sites in which the newly born stars are located \citep[such as the filamentary structures, e.g.][]{Molinari10_special}. In particular the background variability can be significant in FIR Galactic Plane data, especially at longer wavelengths \citep[e.g.,][]{Molinari10_special,Peretto10}, and models must account for its high variations across each source. Furthermore the Galactic Plane is dense and crowded and it is often the case that sources are partially blended together. The problem is further complicated in case of multi-wavelengths study since each band has a different spatial resolution and often single sources are resolved in multiple objects in the high-resolution maps.

Various approaches based on different techniques (e.g. \textit{Cutex}, \citet{Molinari11}; \textit{getsources}, \citet{Men'shchikov12})  have been developed specifically for the new \Her\ FIR data, in particular for the Galactic surveys.  \textit{Cutex} in particular is the standard source extraction algorithm used by the Hi-GAL team. It identifies the compact source in the second derivative image of the sky, and fits the sources with a 2d-Gaussian model. In this approach the physical source diameter, FWHM$_{dec}$, also needed to constraint the source flux, is evaluated through the deconvolution of the HPBW with the FWHM$_{\lambda}$ of the source measured at each band and it is therefore wavelength-dependent \citep[e.g.][]{Molinari11,Elia10,Veneziani12}.

We decide to adopt a different approach and to estimate the flux consistently within the \textit{same volume} of gas and dust at all wavelengths regardless of the different spatial resolutions. A similar approach have been adopted by \citet{Olmi12} to study the ClMF in Hi-GAL fields. 

For this purpose we used a new algorithm, \textit{Hyper} (HYbrid Photometry and Extraction Routine), fully described in \citet{Traficante14_Hyp}. It has been designed specifically to take into account the complexities generated by the new datasets such as the Galactic Plane observations made with \Her, in particular the high background variability and the source crowding. \Hyp\ is based on a hybrid approach between the classical aperture photometry and a 2d-Gaussian modelling of the sources. In the case of multi-wavelength analysis, the region of integration is defined at a particular wavelength and the same region of the sky is used to integrate the flux at all wavelengths. In the following Sections, along with the main results of this work we will describe the \Hyp\ parameters tuned for this source extraction and photometry.

\subsection{The source catalogues}\label{sec:source_identification}
The sources are initially identified at each wavelength separately on a modified high-pass filtered map using in sequence the \texttt{find} and \texttt{gcnrtd} IDL routines \citep{Traficante14_Hyp}. These routines fit a Gaussian model to recognize the peaks above a given threshold \thres\, defined as a multiple of the \textit{r.m.s.} of the filtered map, $\sigma_{f}$. A reference threshold value for each \Her\ wavelength, which minimizes the false positives and maximizes the identification of real sources is provided in \citet{Traficante14_Hyp}. However the source recovery for a given value of \thres\ depends on variations of the (local) background, the source crowding and the cloud size, which differ from cloud to cloud (see also the discussion in Section \ref{sec:flux_completeness}). We tested several values of \thres\ around the reference values discussed in \citet{Traficante14_Hyp} and after visual inspections of randomly selected IRDCs we set \thres=[6.0,5.0,4.0,4.0]$\cdot\sigma_{f}$ at [70,160,250,350] \mum\ respectively. These thresholds are a conservative compromise between source recovery and false identifications. From our visual inspections, less than 1$\%$ of the sources appear as false positives. The completeness of our catalogue based on this extraction is discussed in Section \ref{sec:flux_completeness}.

\begin{figure}
\centering
\includegraphics[width=9cm]{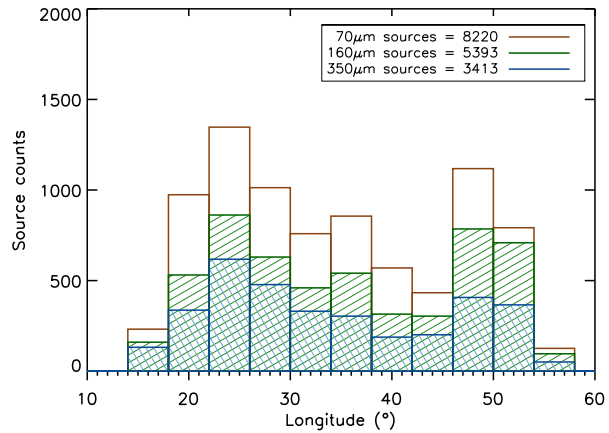}
\caption{Longitude distribution of sources independently identified at 70 (brown), 160 (green) and 350 (blue) \mum. The distribution of 250 \mum\  sources is very similar to the 160 \mum\ distribution and it is therefore not shown.}
\label{fig:sources_all_wave_longitude_distribution}
\end{figure}

From this extraction we produced a catalogue of sources independently identified at each wavelength. \Hyp\ identified 8220, 5393, 4967, 3413 sources at 70, 160, 250 and 350 \mum\ respectively. They are associated with 2070, 1640, 1621, 1246 IRDCs, respectively $\simeq59\%$, $\simeq47\%$, $\simeq46\%$ and $\simeq36\%$ of the whole sample. The high number of 70 \mum\ sources compared to the other wavelengths is likely due to the high spatial resolution of the 70 \mum\ data, which allows the resolution of close sources possibly unresolved at longer wavelengths. Conversely, the low spatial resolution of the 350 \mum\ band blends sources resolved at shorter wavelengths. The longitude distribution of sources at 70, 160 and 350 \mum\ is in Figure \ref{fig:sources_all_wave_longitude_distribution}. The 250 \mum\ source distribution is very similar to that of the 160\,\mum\ sources and is not showed.

The multi-wavelength catalogues contain the photometry of the sources observed at all the specified wavelengths, as discussed in the next Sections, starting from a reference wavelength used to identified the candidates. The reference wavelength is 160 \mum, the highest resolution wavelength that is in common among protostellar and starless clumps.
We initially generated a merged catalogue of 160, 250 and 350 \mum\ sources to produce a list of starless and protostellar clump candidates in our selection of IRDCs. At the location of each 160 \mum\ source, a 250 and/or 350 \mum\ counterpart is associated if a source is identified within a radius equal to half the 160 \mum\ beam, 6.7\arcsec. The merged catalogue contains 1723 clumps associated with 764 different IRDCs ($\simeq22\%$).

A fraction of IRDCs identified in Paper I may not have a counterpart in the \Her\ wavelengths and may not be real clouds \citep[e.g.][]{Wilcock12}, however the small fraction of IRDCs with identified \Her\ compact sources here is due to a combination of factors. First, the merged catalogue does not consider sources detected at only one or two wavelengths, nor objects detected at 160, 250 and 350 \mum\ but with centroids separated by more than 6.7\arcsec\ from the 160 \mum\ centroid. Second, most IRDCs smaller than $\simeq20$\arcsec\ are too small to be associated with a compact \Her\ sources. Note that in particular with this choice we are not including in the catalogues the sources observed at 250 and 350 \mum\ with no counterparts at 160 \mum. We identified $\simeq400$ sources observed at 250 and 350 \mum\ only. Some of these sources are potentially extremely young, their envelope so cold and/or diffuse that they do not emit above the background at wavelengths $\lambda\leq160$ \mum. However, we cannot extract physical parameters for sources identified at only two wavelengths, therefore we restrict the analysis to the clumps already visible at 160 \mum. A multi-wavelength analysis of these very cold clumps is the subject of a subsequent work.

\begin{figure}
\centering
\includegraphics[width=9cm]{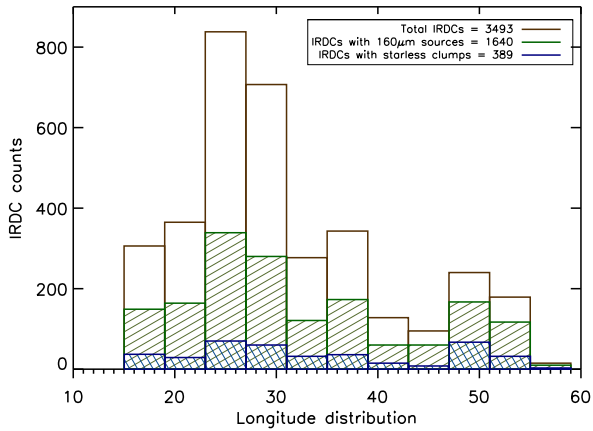}
\caption{Longitude distribution in the range $15\grad\leq l\leq55\grad$ of IRDCs with at least one starless clump (389, blue) together with the distribution of the IRDCs with at least one 160 \mum\ detection (1640, green) and the distribution of the total number of IRDCs in our catalogue with known distances (3493, brown).} 
\label{fig:IRDC_starless_longitude_distribution}
\end{figure}

\begin{figure}
\centering
\includegraphics[width=9cm]{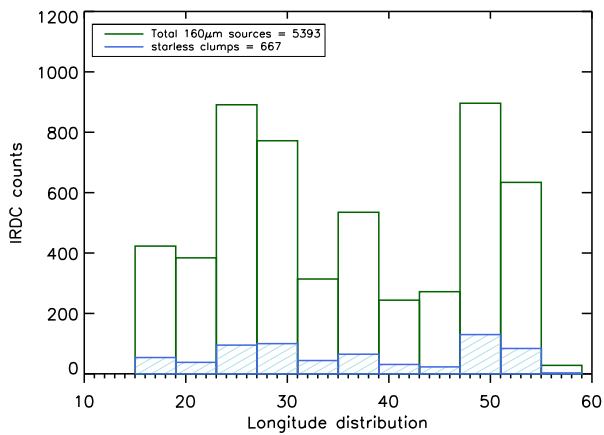}
\caption{Longitude distribution of sources. In green the 160\mum\ sources (5393) and in azure the starless clumps (667).}
\label{fig:source_starless_longitude_distribution}
\end{figure}

\begin{figure}
\centering
\includegraphics[width=9cm]{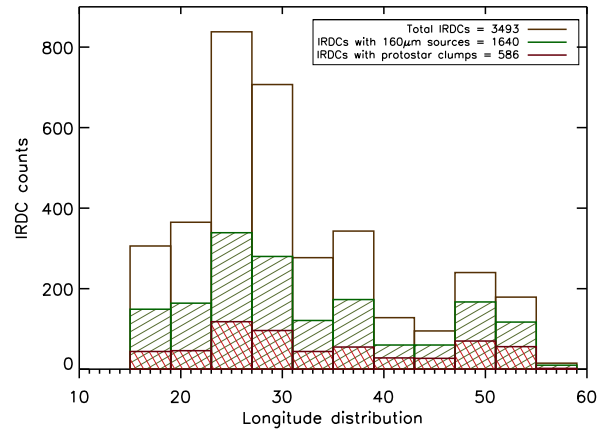}
\caption{IRDCs longitude distribution of IRDCs with at least one protostellar clump (586, red). The other bars are as shown in  Figure \ref{fig:IRDC_starless_longitude_distribution}.}
\label{fig:IRDC_protostar_longitude_distribution}
\end{figure}

\begin{figure}
\centering
\includegraphics[width=9cm]{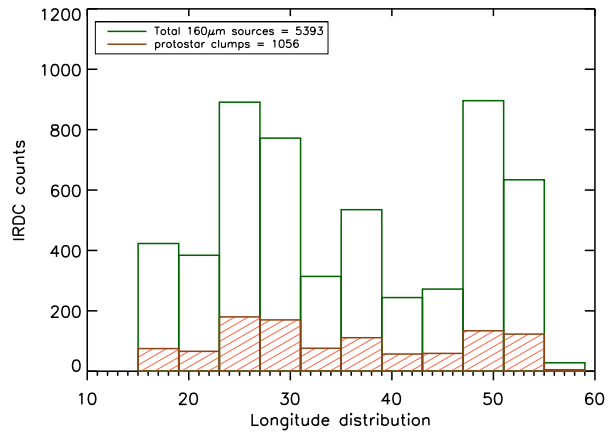}
\caption{Longitude distribution of protostellar clumps in light red (1056). The other bar is as shown in  Figure \ref{fig:source_starless_longitude_distribution}.}
\label{fig:source_protostar_longitude_distribution}
\end{figure}

  
An independent extraction run used to produce a list of 70 \mum\ sources has been combined with the previous sample to produce two final catalogues:

\begin{itemize}
\item[1.] \textit{Sources identified at 160, 250 and 350 \mum\ but without counterparts at 70 \mum}. These sources are classified as starless clumps. The catalogue contains 667 clumps associated with 389 IRDCs. The IRDCs and source longitude distributions are shown in Figures \ref{fig:IRDC_starless_longitude_distribution} and \ref{fig:source_starless_longitude_distribution} respectively. 

\item[2.]  \textit{Sources identified at 160 , 250 and 350 \mum\ with (at least) one counterpart at 70 \mum}. A 70 \mum\ source is associated with each clump if its distance from the centroid of the 160\mum\ source is less or equal to half the 160 \mum\ beam, 6.7\arcsec. These sources are classified as protostellar clumps. This catalogue contains 1056 sources associated with 586 IRDCs. 
The IRDC and source longitude distributions are shown in \ref{fig:IRDC_protostar_longitude_distribution} and \ref{fig:source_protostar_longitude_distribution} respectively. 

\end{itemize}

In total 211 clouds ($\simeq6\%$ of the total) contain both starless and protostellar clumps. 

\begin{figure}
\centering
\includegraphics[width=9cm]{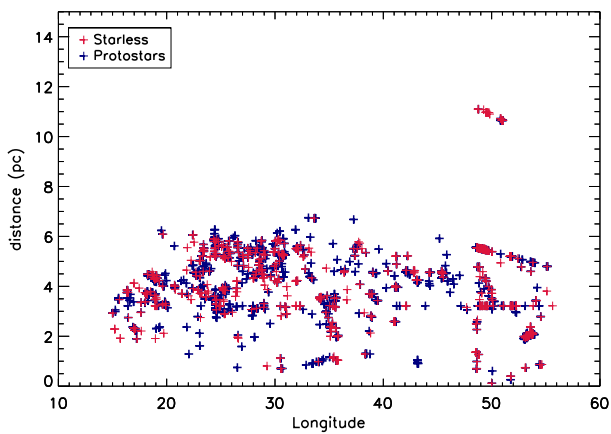}
\caption{Starless and protostellar clump longitude distribution as function of their parent IRDC distance. The two distributions look similar and there are no regions with a clear predominance of starless or protostellar clumps. The majority of the sources are located around $25\grad\leq l \leq30\grad$ and $l=50\grad$. The median distances of the two distributions are $d=4.21$ kpc and $d=4.23$ kpc for starless and protostellar clumps respectively.}
\label{fig:IRDC_tot_distance_distribution}
\end{figure}

Figures 1--5 show that the majority of the sources are located in the longitude range $25\grad\leq l \leq30\grad$, and a second peak in the source distribution occurs around $l=50\grad$, for both starless and protostellar clumps.  The longitude distribution of the clumps as a function of their distance is shown in Figure \ref{fig:IRDC_tot_distance_distribution}. The clump distances are assumed equal to the distance of their parent IRDCs. The median distances are $d=4.2$ kpc for both the  starless and protostellar clumps. 

The distribution of clumps is sparse across the Galaxy with two LOS which contain the majority of the clumps and of the 160 \mum\ sources. The region around $25\grad\leq l \leq30\grad$ corresponds to sources mostly located at a distance $d\simeq5$ kpc, which is where the LOS passes the tangent point across the Scutum-Crux arm (following the \citet{Russeil03} spiral Galaxy model). Being close to the tangent point, the distance ambiguity is minimal for these sources and there are likely to be associated with the Scutum-Crux arm. The source distribution along this LOS is not dominated by starless or protostellar clumps. The peak around $l=50\grad$ corresponds to the tangent point of the Sagittarius-Carina arm, but passes also through the Perseus and the Norma-Cygnus arms. Some of the clumps in this region come from the furthest IRDCs present in our catalogue (see Figure \ref{fig:IRDC_tot_distance_distribution}), located at $\sim10$ kpc and are likely to be associated with the Perseus arm. The peak at $l=50\grad$  in the source distribution is more evident in the starless clumps. Although some 70 \mum\ sources could be too faint to be observed in the furthest clouds, the mean IRDCs distance around  $l=50\grad$ is similar to the mean distance of the clouds around  $l=30\grad$, 5 kpc and 4.6 kpc respectively. Therefore, the peak at $l=50\grad$ in the starless distribution is likely to be a sign of younger star-forming region along this LOS compared to the Scutum-Crux region. The distribution in Figure \ref{fig:IRDC_tot_distance_distribution} indicates that the Hi-GAL sources are mainly located on the Galactic arms, although some are in  interarms regions, as already noted by \citet{Russeil11}.

\subsection{Photometry}\label{sec:photometry}

The flux  of each source is estimated by integrating the emission within the \textit{same} elliptical region at all wavelengths. The semi-axes $a$ and $b$ of the elliptical aperture are equal to the FWHM$_{G}$ of the 2d-Gaussian fit to each source estimated at a reference wavelength, which can be different from the wavelength used to initially identify the clumps \citep{Traficante14_Hyp}. For the analysis here we set the reference wavelength to ${\lambda}=250$ \mum. The FWHM$_{G}$ of each fit can vary from a minimum of $1\cdot\mathrm{FWHM}_{\lambda}$ (18$\arcsec$ at ${\lambda}=250$ \mum) up to $2\cdot\mathrm{FWHM}_{\lambda}$ (36$\arcsec$). This size  at ${\lambda}=250$ \mum\ encompasses a region of at least $1.5\cdot\mathrm{FWHM}_{\lambda}$ at ${\lambda}=350$ \mum\ within the integration area. The range of allowed FWHM$_{G}$ has been chosen in order to account for point-like and slightly elongated compact sources and, at the same time, to avoid highly elongated fits likely contaminated by the underlying filaments aligned with some of the sources. The profile at ${\lambda}=250$ \mum\ also defines the source size as described in Section \ref{sec:rad_temp_mass}.

We note that within each elliptical aperture the clump could be resolved in multiple objects at 70 and 160\,\mum. In these cases, the closest source in the cluster is identified as the counterpart and we assign  each clump the 70 and 160,\mum\ flux arising from all the sources within the elliptical integration area. This choice is consistent with our approach of evaluating the flux within the \textit{same} area, since the sources are blended within the beam at longer wavelengths and they all contribute to the observed emission. In the case of these resolved clusters, a specific keyword in the catalogue (``cluster") indicates the number of multiple sources observed at 70 and 160\,\mum\ within each integration area.

The source flux evaluation is done at each wavelength after the background emission is subtracted and, in case of blended sources, the flux arising from the companions is removed.
 \Hyp\ evaluates the local background by selecting various square regions across each source and modelling the emission in each region with polynomial functions of different orders (from the zeroth up to the fourth). The fit which produces the lowest \rms\ of the residuals is assumed as best fit and subtracted as background contribution. 

Two (or more) sources are considered blended if their centroids are closer than a fixed distance. This distance is automatically evaluated by the algorithm and it is equal to twice the maximum allowed FWHM$_{G}$ (e.g. 72$\arcsec$), the maximum distance at which two elliptical apertures can be partially overlapped. Each source and its blended companions are fitted simultaneously with a multi-Gaussian function using the \texttt{mpfit} IDL routine \citep{Markwardt09}. The parameters of the fit are used to build 2d-Gaussian models of the companions which are then subtracted from the image. The flux of the companion-subtracted source is then evaluated as in the isolated source case. 
A detailed description of the background subtraction and source de-blending strategies can be found in \citet{Traficante14_Hyp}. 
However the source fluxes still require to be corrected for two factors: the aperture and colour corrections.

\subsubsection{Aperture corrections}\label{sec:aperture_corr}

Aperture corrections A$_\mathrm{c}$ are needed in aperture photometry to compensate for the flux distributed outside the integration area due to the convolution of the true sky signal with the instrumental beam, the PSF response. These corrections therefore depend on the beam at each different wavelength and, for point-like sources, can be estimated by measuring the source emission of reference sources with known flux. A further correction factor has to be applied for extended sources. 

For PACS, the aperture corrections are known for circular apertures as function of aperture radius\footnote{$\mathrm{http://herschel.esac.esa.int/twiki/pub/Public/PacsCalibrationWeb/} \\ \mathrm{pacs\_bolo\_fluxcal\_report\_v1.pdf}$}. The equivalent aperture radius $r_{eq}$ of each source, described in Section \ref{sec:rad_temp_mass}, is used to estimate the corresponding A$_\mathrm{c}$.

For SPIRE, aperture corrections are known for circular apertures as function of aperture radius and source spectral index\footnote{$\mathrm{http://herschel.esac.esa.int/twiki/pub/Public/PacsCalibrationWeb/} \\ \mathrm{http://herschel.esac.esa.int/Docs/SPIRE/spire\_handbook.pdf}$}. The corrections are estimated at fixed aperture radii of 22\arcsec, 30\arcsec\ and 40\arcsec\ at 250, 350 and 500 \mum\ respectively. Our aperture radius depends the 2d-Gaussian fit at $\lambda=250$\,\mum\ to each source and it is fixed at each wavelength (Section \ref{sec:photometry}). However, the range of allowed FWHM$_{G}$ is from 18\arcsec\ to 36\arcsec, therefore we decided to adopt the proposed corrections without further assumptions and assuming a spectral index of $\beta=2.0$ (as discussed in Section \ref{sec:mass_luminosity}).

\subsubsection{Colour corrections}\label{sec:colour_corr}

\begin{figure}
\centering
\includegraphics[width=9cm]{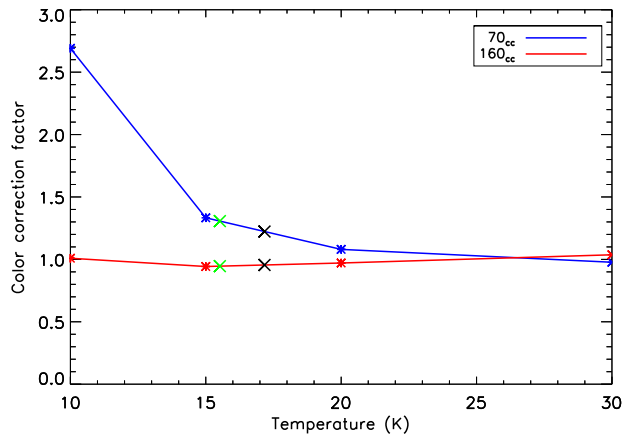}
\caption{PACS colour correction curves for both the 70 and 160 \mum\ filters, extrapolated from the colour correction values at T=10, 15, 20 and 30 K described in the PICC-ME-TN-038 PACS report. The green and black crosses show the colour corrections applied for the starless and protostellar clumps flux distributions respectively. Their values are 1.307 and 0.946 for starless and 1.223 and 0.955 for protostellar clumps at 70 and 160 \mum\ respectively, corresponding to a temperature of T=15.52 K and 17.17 K for starless and protostellar clumps respectively.}
\label{fig:PACS_col_corr}
\end{figure}

Colour corrections are needed to convert the  flux density measured by PACS and SPIRE into the monochromatic flux density for each observed object, which depends on the intrinsic temperature and spectral energy distribution (SED) of the source. The colour corrections are available online for both PACS\footnote{$\mathrm{http://herschel.esac.esa.int/twiki/pub/Public/} \\ \mathrm{PacsCalibrationWeb/cc\_report\_v1.pdf}$} and SPIRE\footnote{$\mathrm{http://herschel.esac.esa.int/twiki/pub/Public/PacsCalibrationWeb/} \\ \mathrm{http://herschel.esac.esa.int/Docs/SPIRE/spire\_handbook.pdf}$} instruments. The colour corrections have been measured for different spectral indexes and blackbody (and greybody) temperatures for both PACS and SPIRE instruments. We adopted a fixed spectral index $\beta=2.0$ (see Section \ref{sec:mass_luminosity}).
 The PACS colour correction curves extrapolated for temperatures in the range $10\leq\mathrm{T}\leq30$ K for both the  PACS 70 and 160 \mum\ filters are shown in Figure \ref{fig:PACS_col_corr}. In order to extrapolate the colour corrections for our sample however we need to know the greybody temperature in advance and this poses a circular problem \citep{Pezzuto12}. So we estimate the colour corrections following an iterative procedure. We first fixed the colour corrections assuming T=15 K for both starless and protostellar clumps, in agreement with previous estimations of their envelope temperature \citep[e.g.][]{Veneziani12}. We obtained a first estimation of the greybody temperatures of our sources resulting in median temperature values of T=14.37 K and T=15.84 K for starless and protostellar clumps respectively.  These new values were then used to estimate the first iteration of the colour corrections from the curve in Figure \ref{fig:PACS_col_corr}. The SEDs obtained with these new corrections give median temperature values of T=15.52 K and 17.17 K for starless and protostellar clumps respectively. Adopting the colour corrections for these temperatures resulted in 
 median temperatures of T=15.49 K and T=17.13 K for starless and protostellar clumps respectively, sufficiently 
 close to the previous estimate, that no further iterate was carried out. The adopted colour corrections are 1.307  and 0.946 for starless and 1.223 and 0.955 for protostellar clumps at 70 and 160 \mum\ respectively. 
The SPIRE colour corrections do not vary significantly in the range 10-20 K at $\beta=2.0$. Therefore, we adopted the publicly available values assuming T=15 K for both starless and protostellar clumps. The values are 1.223 and 0.955 for the 250 and 350 \mum\ fluxes respectively.

\subsubsection{Flux error}
The quoted error for the flux in the PACS and SPIRE reference manuals is, after aperture and colour corrections, $<5\%$. This error has to be combined with the error associated with the \Hyp\ photometry. The error quoted in the catalogues is the flux error estimated as the local sky r.m.s. per pixel multiplied for the number of pixels in the area over which the source flux is integrated \citep{Traficante14_Hyp}. In addition, the precision in recovering the source flux depends on the intensity of each source and the local background variability. For each source we assume a conservative error of $5\%$ of its flux associated with the aperture and colour corrections plus an error of $15\%$ from \Hyp\ measurements. Hence each source flux has associated a conservative error of $20\%$ of its flux at every wavelength.

\subsection{The clumps catalogues}
\begin{figure*}
\centering
\includegraphics[width=10cm]{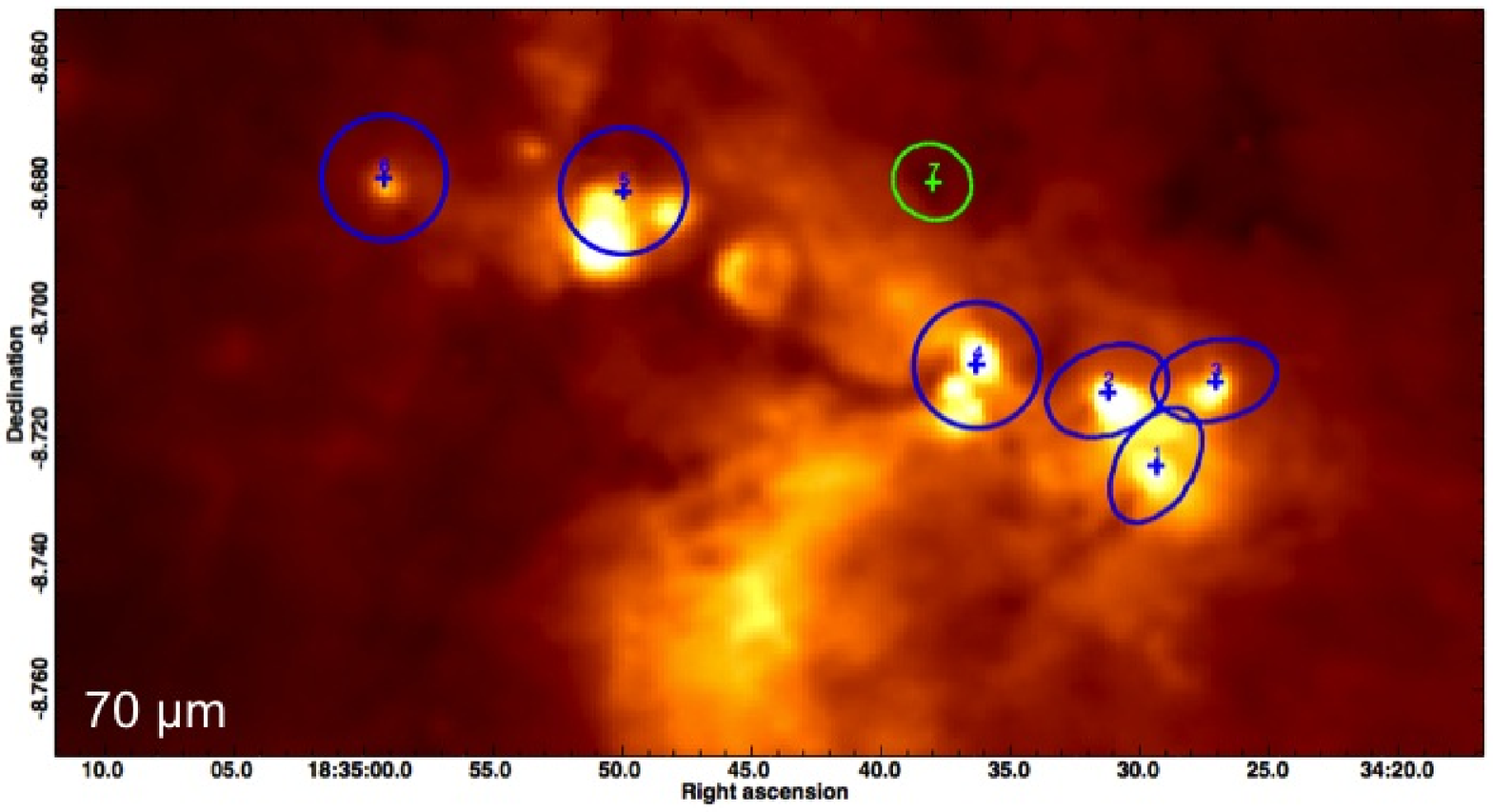}\\  
\includegraphics[width=10cm]{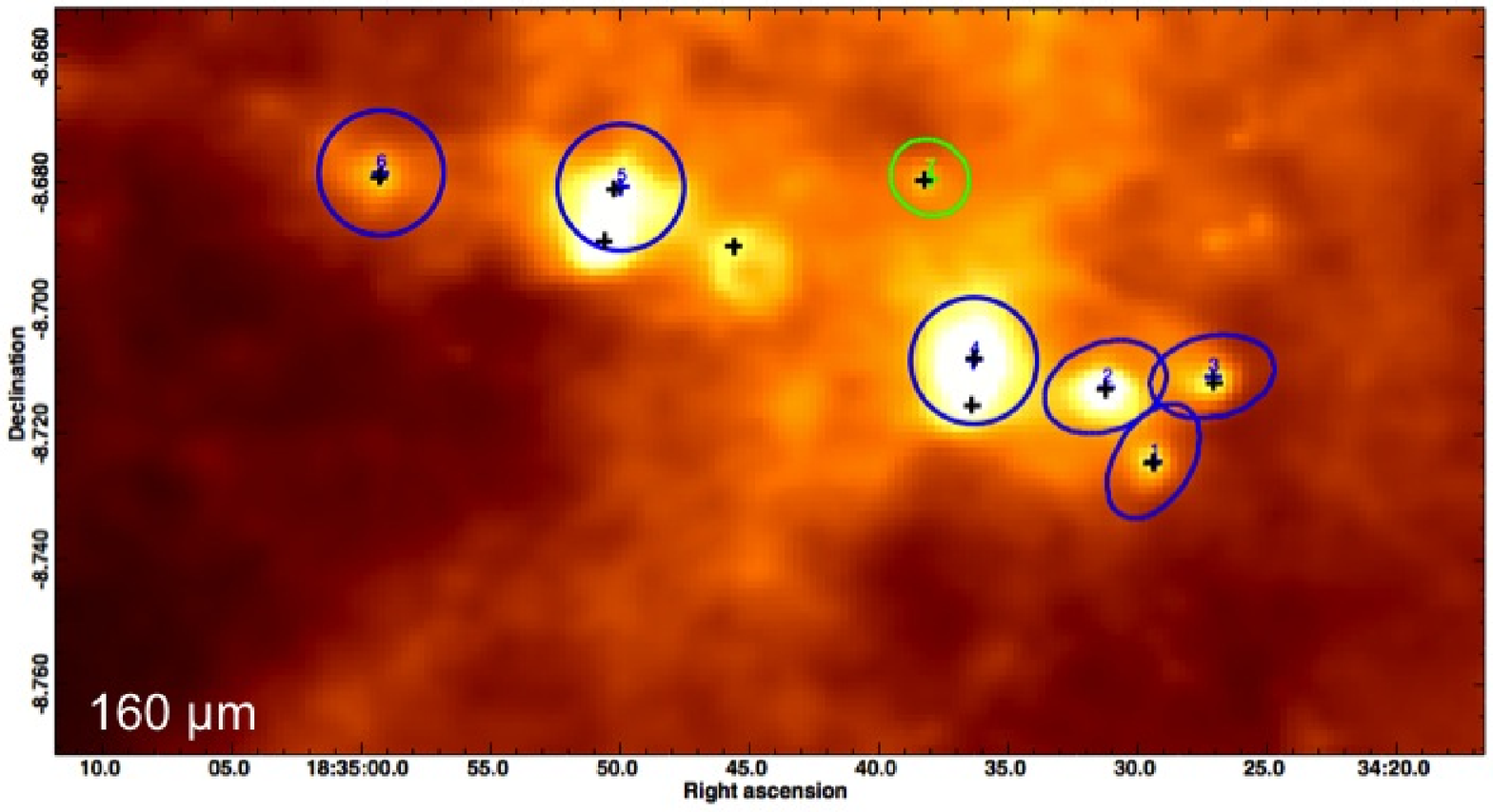}\\
\includegraphics[width=10cm]{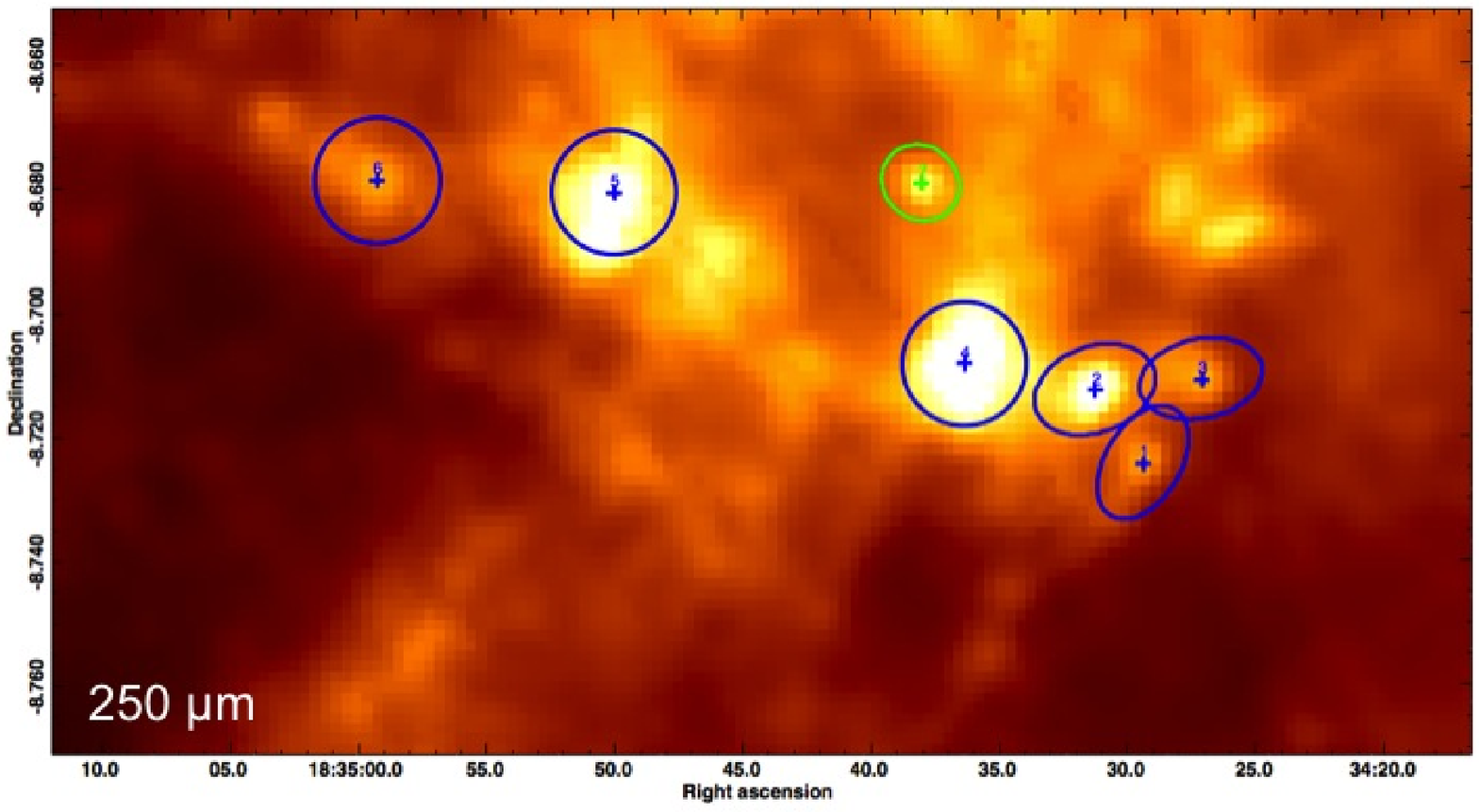}\\
\includegraphics[width=10cm]{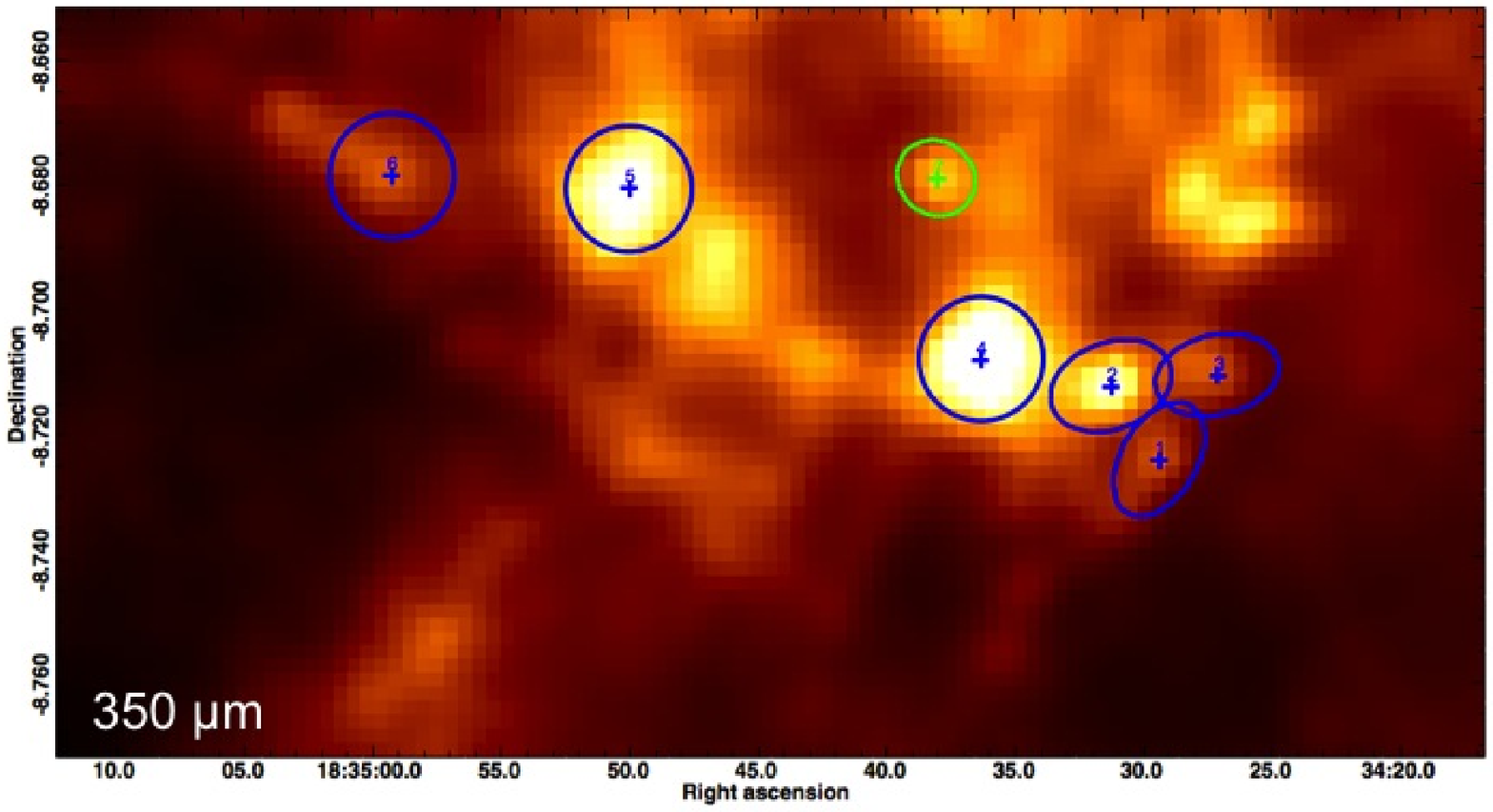}\\
\caption{The FIR counterpart of the IRDC catalogued as HGL23.271-0.263. From top to the bottom: 70, 160, 250 and 350 \mum\ images. The protostellar clumps are shown in blue and the starless clumps are shown in green. The ellipses correspond to the elliptical region, defined at 250 \mum, used to estimate the source fluxes at all wavelengths. The black crosses in the 160 \mum\ image shows the sources initially identified in this map.} 
\label{fig:HGL23.271-0.263}
\end{figure*}

An example of source extraction and photometry at the four considered wavelengths in the SDC23.271-0.263, catalogued as HGL23.271-0.263, is shown in Figure \ref{fig:HGL23.271-0.263}. \Hyp\ identified seven sources observed at 160, 250 and 350 \mum\ simultaneously. Six sources have a 70 \mum\ counterpart and are therefore classified as protostellar clumps, one is classified as starless clump. The 2d-Gaussian fits of the clumps evaluated at 250 \mum\ are shown in green and blue for starless and protostellar clumps respectively. The sources identified at 160 \mum\ are indicated as black crosses in the 160 \mum\ map. 

The catalogue parameters obtained for the sources extracted in HGL23.271-0.263 are listed in Table \ref{tab:hyper_output_file}. The quantities reported for each source are: the IRDC name and the source identification number as extracted from \Hyp\ (Columns 1--2), which corresponds to the number in the associated region file; the wavelength (Column 3); the source peak flux in  MJy/sr and Jy, evaluated at the centroid pixel (Columns 4--5); the source integrated flux (not corrected) and its associated error (Columns 6--7); the sky r.m.s. before and after the background subtraction (Columns 8--9); the polynomial order used to estimate the background (Column 10); the source semi-axes and the position angles as obtained from 2d-Gaussian fitting  at the reference wavelength, 250 \mum\ (Columns 11--13); the convergence status of the fit (Column 14). This status is 0 if the fitting routine converges, otherwise it assumes the value $-1$ if the source fit is too elongated (axes ratio $\geq2.5$) or $-2$ if the fitting routine did not converge. In both these cases the code forces the semi-axes to be equal to the average of the minimum and maximum FWHM possible values (i.e., $1.5\cdot\mathrm{FWHM}_{250\mu m}$); the position of the source centroids at each different wavelength, in both Galactic and Equatorial (J2000) coordinates  (Columns 15--18); the distance between the 160 \mum\ centroid and the centroids of the source counterparts at the other wavelengths (Column 19); the number of sources with their 2d-Gaussian profile overlapped and de-blended before the source flux integration (Column 20). The number of multiple sources resolved at wavelengths other than the 250 \mum\ within the integration area (Column 21).

\begin{landscape}
\begin{table}
\centering
\begin{tabular}{c|c|c|c|c|c|c|c|c|c|c|c|c|c|c|c|c|c|c|c|c}
\hline
\hline
map & sou. & band & peak fl. & peak fl. & flux & err${\_}$flux & sky${\_}$nob. & sky & ord. & fwhm & fwhm  & PA & status & glon & glat &  ra & dec & dist. & deb. & clust\\
& & (\mum) & (MJy/sr) & (Jy) & (Jy) & (Jy) & (Jy) & (Jy) &  & ($\arcsec$) & ($\arcsec$) & ($^{\circ}$) & & ($^{\circ}$) & ($^{\circ}$) & ($^{\circ}$) & ($^{\circ}$) &  ($\arcsec$) &  &\\
\hline

 23.271-0.263  &       1  &       70  &      2249.96  &         0.54  &        16.60  &         2.21  &         0.213  &          0.142  &          3  &        21.91  &         36.07  &        149.14  &          0  &          23.242  &          -0.240  &         278.623  &          -8.726  &           3.35  &          2  &          1  \\
 23.271-0.263  &       1  &      160  &      2623.40  &         1.25  &        37.29  &         3.12  &         0.474  &          0.282  &          1  &        21.91  &         36.07  &        149.14  &          0  &          23.243  &          -0.239  &         278.623  &          -8.725  &           0.00  &          2  &          1  \\
 23.271-0.263  &       1  &      250  &      1102.27  &         0.93  &        24.60  &         4.16  &         0.734  &          0.501  &          1  &        21.91  &         36.07  &        149.14  &          0  &          23.243  &          -0.239  &         278.622  &          -8.724  &           1.39  &          2  &          1  \\
 23.271-0.263  &       1  &      350  &       435.34  &         0.65  &        14.70  &         5.08  &         0.949  &          0.815  &          1  &        21.91  &         36.07  &        149.14  &          0  &          23.243  &          -0.240  &         278.623  &          -8.725  &           3.11  &          2  &          1  \\
 23.271-0.263  &       2  &       70  &     12507.02  &         3.01  &       100.25  &         4.41  &         0.297  &          0.265  &          1  &        25.02  &         36.07  &        110.66  &          0  &          23.256  &          -0.240  &         278.630  &          -8.714  &           2.94  &          2  &          1  \\
 23.271-0.263  &       2  &      160  &     10611.79  &         5.05  &       114.53  &         4.45  &         0.520  &          0.377  &          4  &        25.02  &         36.07  &        110.66  &          0  &          23.257  &          -0.240  &         278.630  &          -8.713  &           0.00  &          2  &          1  \\
 23.271-0.263  &       2  &      250  &      4201.20  &         3.56  &        95.12  &         9.93  &         1.295  &          1.119  &          1  &        25.02  &         36.07  &        110.66  &          0  &          23.257  &          -0.240  &         278.630  &          -8.713  &           0.83  &          2  &          1  \\
 23.271-0.263  &       2  &      350  &      1572.63  &         2.37  &        64.42  &         5.86  &         1.031  &          0.881  &          1  &        25.02  &         36.07  &        110.66  &          0  &          23.257  &          -0.240  &         278.630  &          -8.712  &           1.75  &          2  &          1  \\
 23.271-0.263  &       3  &       70  &      3661.37  &         0.88  &        43.39  &         3.02  &         0.241  &          0.189  &          1  &        23.10  &         36.07  &        102.28  &          0  &          23.250  &          -0.225  &         278.613  &          -8.712  &           2.16  &          2  &          1  \\
 23.271-0.263  &       3  &      160  &      3240.27  &         1.54  &        70.44  &         4.08  &         0.345  &          0.359  &          1  &        23.10  &         36.07  &        102.28  &          0  &          23.250  &          -0.225  &         278.613  &          -8.712  &           0.00  &          2  &          1  \\
 23.271-0.263  &       3  &      250  &      1454.45  &         1.23  &        49.61  &         2.31  &         0.399  &          0.271  &          3  &        23.10  &         36.07  &        102.28  &          0  &          23.251  &          -0.224  &         278.613  &          -8.711  &           2.97  &          2  &          1  \\
 23.271-0.263  &       3  &      350  &       375.82  &         0.56  &        14.01  &         2.41  &         0.546  &          0.376  &          1  &        23.10  &         36.07  &        102.28  &          0  &          23.252  &          -0.225  &         278.614  &          -8.710  &           8.79  &          2  &          1  \\
 23.271-0.263  &       4  &       70  &      6063.55  &         1.46  &        89.55  &         5.59  &         0.283  &          0.280  &          2  &        36.07  &         36.07  &        252.40  &          0  &          23.271  &          -0.256  &         278.651  &          -8.707  &           2.39  &          0  &          4  \\
 23.271-0.263  &       4  &      160  &     13647.29  &         6.50  &       394.78  &        10.69  &         0.767  &          0.753  &          4  &        36.07  &         36.07  &        252.40  &          0  &          23.271  &          -0.257  &         278.651  &          -8.708  &           0.00  &          0  &          2  \\
 23.271-0.263  &       4  &      250  &      7821.71  &         6.62  &       298.13  &         9.59  &         0.945  &          0.900  &          3  &        36.07  &         36.07  &        252.40  &          0  &          23.271  &          -0.257  &         278.651  &          -8.708  &           1.27  &          0  &          1  \\
 23.271-0.263  &       4  &      350  &      3162.00  &         4.76  &       138.29  &         4.12  &         0.660  &          0.516  &          3  &        36.07  &         36.07  &        252.40  &          0  &          23.271  &          -0.256  &         278.651  &          -8.708  &           1.46  &          0  &          1  \\
 23.271-0.263  &       5  &       70  &      1848.66  &         0.44  &       132.85  &         7.69  &         0.451  &          0.385  &          1  &        36.07  &         36.07  &        230.89  &          0  &          23.321  &          -0.298  &         278.711  &          -8.682  &           8.72  &          0  &          2  \\
 23.271-0.263  &       5  &      160  &      4595.00  &         2.19  &       243.37  &         8.49  &         0.756  &          0.597  &          3  &        36.07  &         36.07  &        230.89  &          0  &          23.321  &          -0.295  &         278.709  &          -8.681  &           0.00  &          0  &          2  \\
 23.271-0.263  &       5  &      250  &      3167.15  &         2.68  &       161.36  &         6.69  &         0.782  &          0.628  &          4  &        36.07  &         36.07  &        230.89  &          0  &          23.321  &          -0.294  &         278.708  &          -8.681  &           3.63  &          0  &          1  \\
 23.271-0.263  &       5  &      350  &      1553.36  &         2.34  &        93.40  &         4.34  &         0.591  &          0.543  &          1  &        36.07  &         36.07  &        230.89  &          0  &          23.321  &          -0.295  &         278.709  &          -8.681  &           0.86  &          0  &          1  \\
 23.271-0.263  &       6  &       70  &      1610.35  &         0.39  &        21.64  &         1.18  &         0.084  &          0.059  &          4  &        36.07  &         36.07  &        117.66  &          0  &          23.339  &          -0.327  &         278.746  &          -8.680  &           4.49  &          0  &          1  \\
 23.271-0.263  &       6  &      160  &      2004.29  &         0.95  &        47.64  &         2.10  &         0.268  &          0.148  &          4  &        36.07  &         36.07  &        117.66  &          0  &          23.340  &          -0.327  &         278.747  &          -8.679  &           0.00  &          0  &          1  \\
 23.271-0.263  &       6  &      250  &       971.20  &         0.82  &        45.75  &         2.06  &         0.434  &          0.193  &          4  &        36.07  &         36.07  &        117.66  &          0  &          23.340  &          -0.327  &         278.747  &          -8.679  &           2.17  &          0  &          1  \\
 23.271-0.263  &       6  &      350  &       465.66  &         0.70  &        22.41  &         2.02  &         0.418  &          0.252  &          4  &        36.07  &         36.07  &        117.66  &          0  &          23.341  &          -0.328  &         278.748  &          -8.678  &           4.99  &          0  &          1  \\
\\
 23.271-0.263  &       7  &      160  &      1392.66  &         0.66  &        13.40  &         1.02  &         0.243  &          0.116  &          1  &        21.17  &         23.21  &        232.83  &          0  &          23.300  &          -0.251  &         278.659  &          -8.679  &           0.00  &          0  &          1  \\
 23.271-0.263  &       7  &      250  &      1663.01  &         1.41  &        32.07  &         2.93  &         0.449  &          0.447  &          1  &        21.17  &         23.21  &        232.83  &          0  &          23.300  &          -0.250  &         278.658  &          -8.679  &           3.78  &          0  &          1  \\
 23.271-0.263  &       7  &      350  &       755.46  &         1.14  &        22.97  &         2.34  &         0.439  &          0.476  &          1  &        21.17  &         23.21  &        232.83  &          0  &          23.300  &          -0.249  &         278.658  &          -8.679  &           5.28  &          0  &          1  \\

\hline
\end{tabular}
\caption{Output parameters for protostellar and starless clumps extracted in HGL23.271-0.263. The catalogues for protostellar (sources 1--6) and starless clumps (source 7) are separated by a blank line in this table. Column 1: IRDC name. Column 2: \Hyp\ source
  number. Column 3: wavelength. Columns 4--5: source peak flux in MJy/sr and Jy. Columns 6--7: source
  integrated flux and source flux error in Jy. This source flux is not corrected for aperture or colour corrections. The flux error is estimated from the local sky
  r.m.s. multiplied by the number of pixels in the area
  over which the source flux is integrated. Column 8--9: r.m.s. of the sky
  evaluated in the rectangular region used to model the background before and
  after the background subtraction respectively. Column 10: polynomial order
  used to model the background. Columns 11--12: minor and maximum FWHM of the 2d-Gaussian fit. Column 13: Source Position Angle. Column 14: goodness of the
  2d-Gaussian fit. Status is equal to 0 if the fit has regularly converged. Columns 15--18: source centroids
  Galactic and Equatorial coordinates (J2000). Column 19: distance from the source centroids in the reference wavelength and the source counterparts at the other wavelengths. Column 20: number of sources identified as companions and de-blended. Column 21: number of source counterparts at each wavelength. It normally is equal to 1 but it can be higher if the source is resolved in more than one counterpart in the high resolution maps.}
\label{tab:hyper_output_file}
\end{table}
\end{landscape}

\section{Starless and protostellar clumps analysis}\label{sec:mass_luminosity} 
The source photometric properties and the source distances are combined to build the SEDs and estimate equivalent radius, temperature, mass and FIR luminosity. We assume a single temperature greybody model with a fixed spectral index to describe the emission of the cold dust envelope of both starless and protostellar clumps. We model the flux of each source $S_{\nu}$ at frequency $\nu$ as \citep[e.g.,][]{Elia10,Giannini12}:

\begin{equation}\label{eq:sed}
S_{\nu}=\frac{M\kappa_{0}}{d^{2}}\bigg(\frac{\nu}{\nu_{0}}\bigg)^{\beta}B_{\nu}(T)\Omega 
\end{equation}
where $\nu_{0}$ and $\kappa_{0}$ are respectively a normalisation frequency and the dust mass absorption coefficient evaluated at  $\nu_{0}$ \citep[$\nu_{0}$=230 GHz and $\kappa_{0}$=0.005 g cm$^{2}$ with a gas to dust ratio of 100,][]{Preibisch93}. The spectral index $\beta$ has been fixed to $\beta=2.0$, in agreement with the standard value for cold dust emission \citep[][]{Hildebrand83} and with previous studies of starless and protostellar clumps \citep[e.g.][]{Motte10,Konives10,Veneziani12}. The total mass (gas+dust) is $M$ and $\Omega$ is the solid angle in which the flux has been evaluated. $B_{\nu}(T)$ is the Planck blackbody function at frequency $\nu$ and temperature $T$. Finally, $d$ is the distance of the source.

For both starless and protostellar clumps we used the fluxes at 160, 250 and 350 \mum\ to model the SED and estimate mass and dust temperature. The FIR luminosity is evaluated by integrating the flux in the range $70\leq\lambda\leq500$ \mum. The 500 \mum\ is extrapolated from the SED fit in both the distributions, where the 70 \mum\ flux is extrapolated from the fit for starless clumps and is obtained from the measured flux for protostellar clumps. The 70 \mum\ flux (and that at shorter wavelengths) of the protostellar clumps is partly influenced by the warm core(s) in the centre \citep[e.g.,][]{Dunham08,Motte10}. A single temperature greybody model cannot account for this second, warm component. Therefore, we do not include the 70 \mum\ emission to estimate the protostellar clump dust temperature and mass.

We note that the temperature gradients across the whole IRDCs can be up to $\simeq 10$K \citep{Peretto10}, but it is negligible across the starless clumps. Furthermore, the differences in the starless clumps mass estimation assuming a single-temperature greybody model or with a detailed radiative transfer model is negligible \citep{Wilcock11}. At the same time, the central regions of the protostellar clumps partially warmed-up by the central core(s) are not included in the model, therefore the model underestimates the protostellar clumps central temperature. As a consequence of Equation \ref{eq:sed}, the model could overestimate the protostellar masses (see in Section \ref{sec:rad_temp_mass}). The adopted single clump-averaged temperature model is however a reasonable approximation to describe the emission arising from the cold, extended dust envelope of both types of objects. A more complete analysis of protostellar properties requires to model at least two different temperature components, using more sophisticated models \citep[e.g.][]{Robitaille07} and using shorter wavelengths to constraint the warm component \citep[e.g.][]{Motte10}. 

The best fit for the greybody model is evaluated with a chi-square minimisation analysis using the \texttt{mpfit} IDL routine \citep{Markwardt09}. In order to restrict the analysis to the sources with good temperature, mass and luminosity estimates, only sources with fits with  $\chi^{2}\leq10$ are considered further. To further delimit the sample, we excluded few sources with $\chi^{2}\leq10$ but with unrealistic temperature values (T$\geq40$ K), likely a consequence of constraining the SED with only three points. The majority of the sources have good fits however, and the final catalogue contains 649 out of a total of 667 starless clumps and 1030 out of 1056 protostellar clumps. 

\begin{figure*}
\centering
\includegraphics[width=8cm]{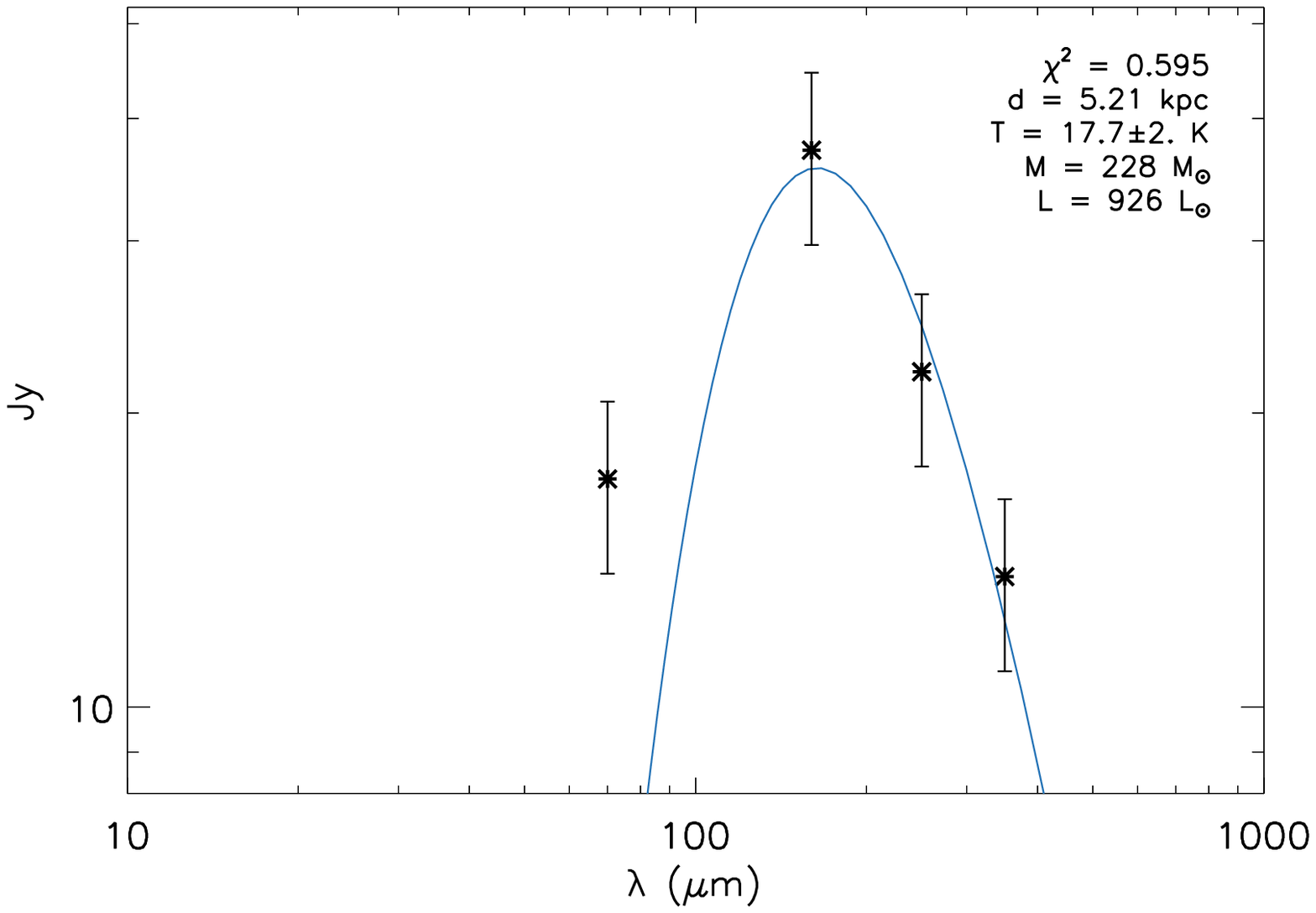}
\includegraphics[width=8cm]{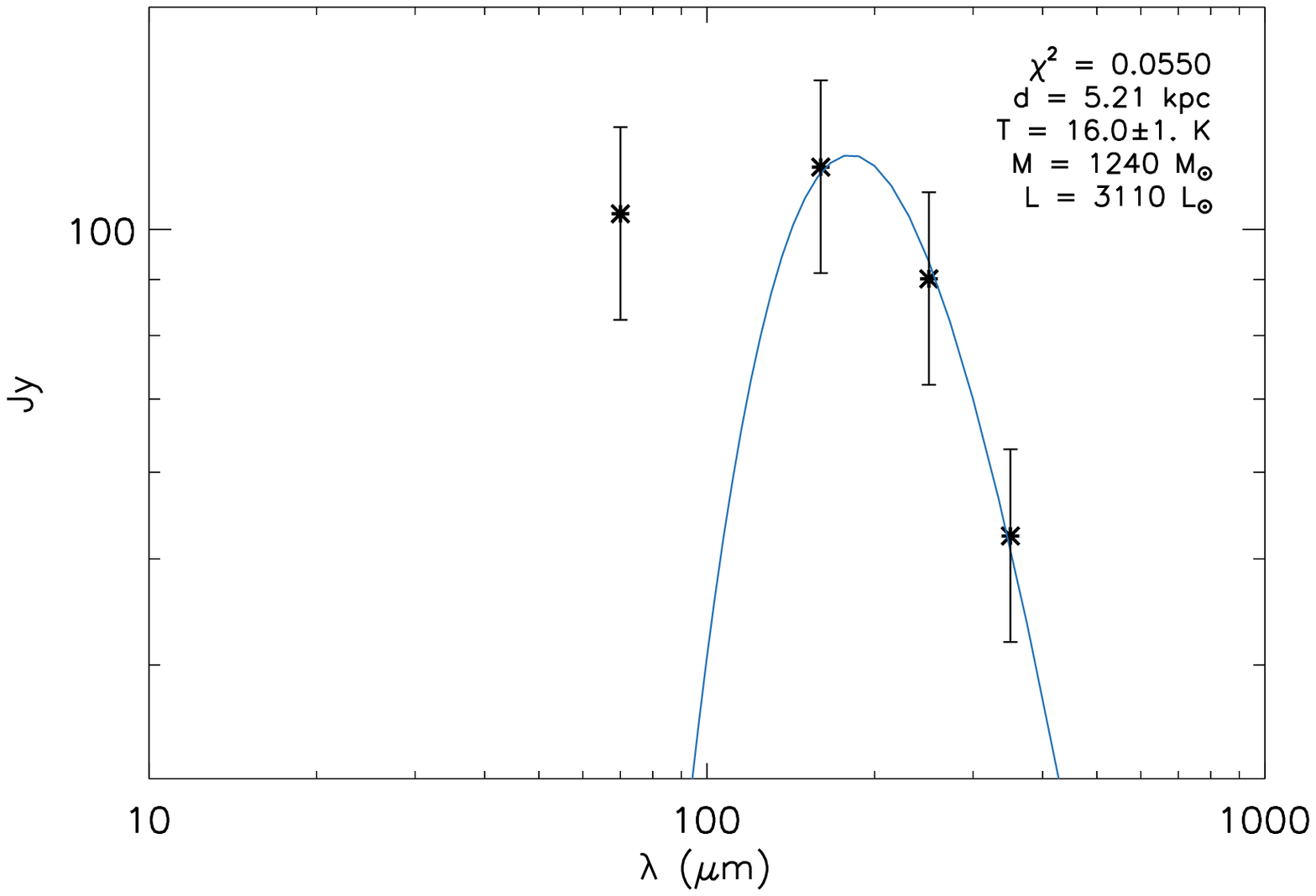}
\includegraphics[width=8cm]{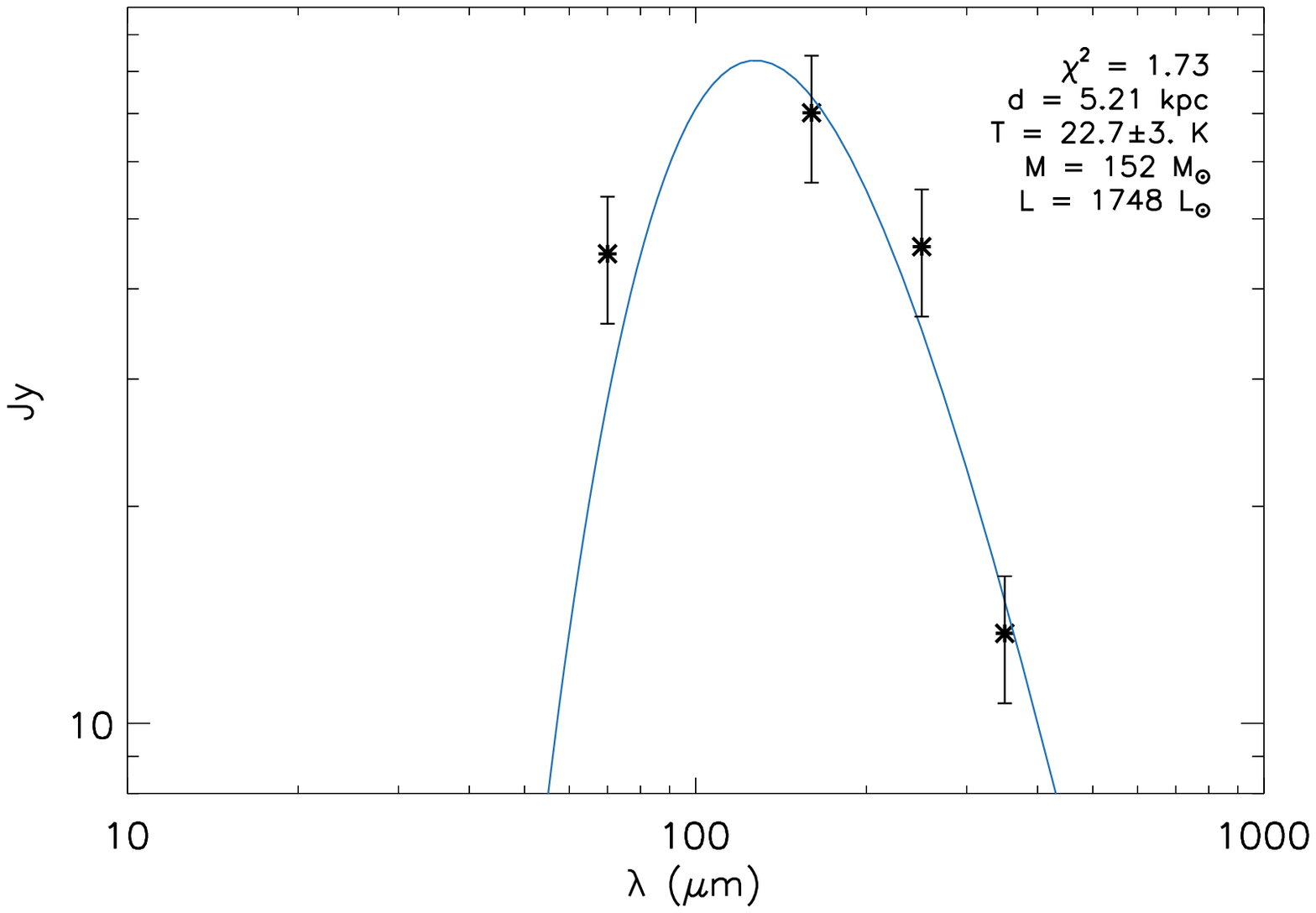}
\includegraphics[width=8cm]{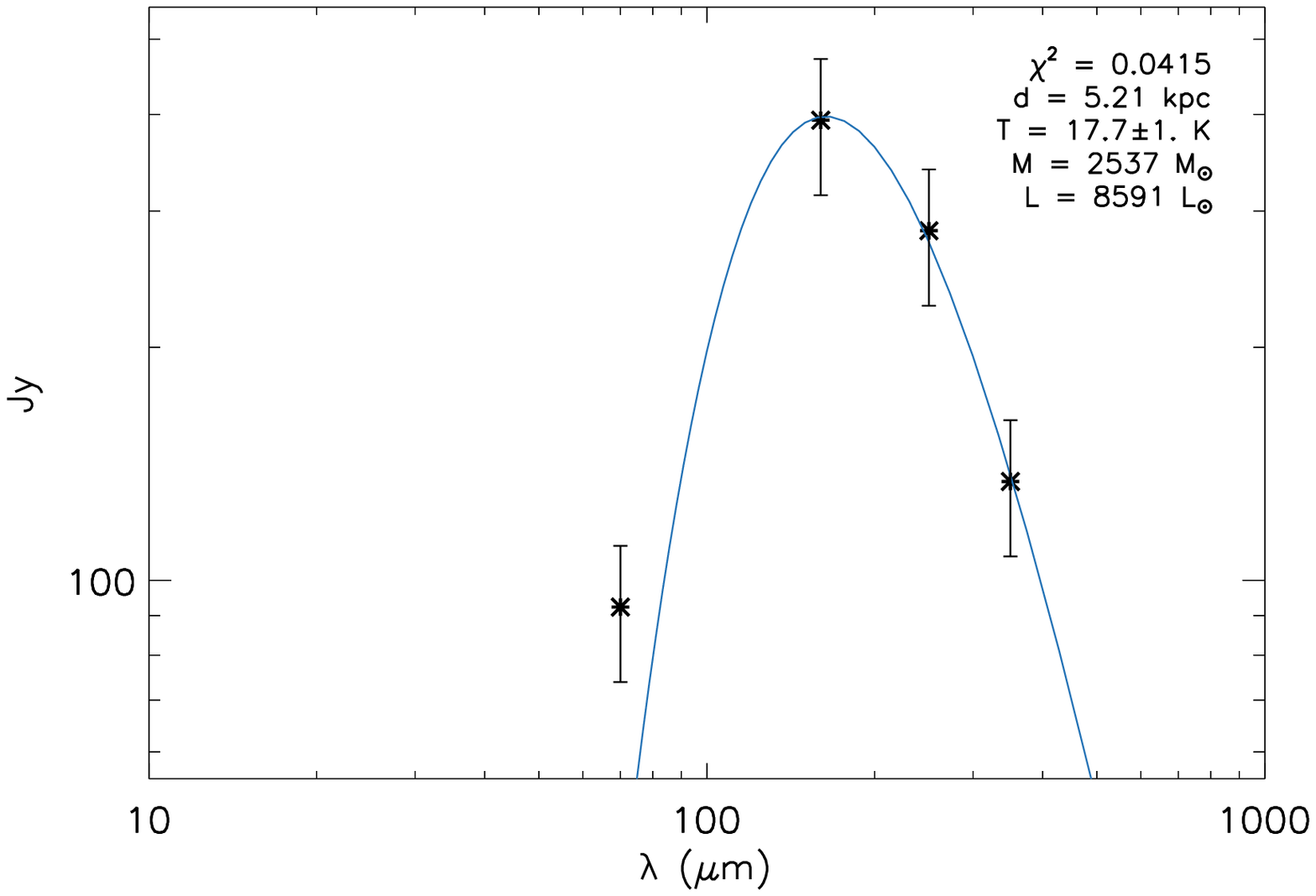}
\includegraphics[width=8cm]{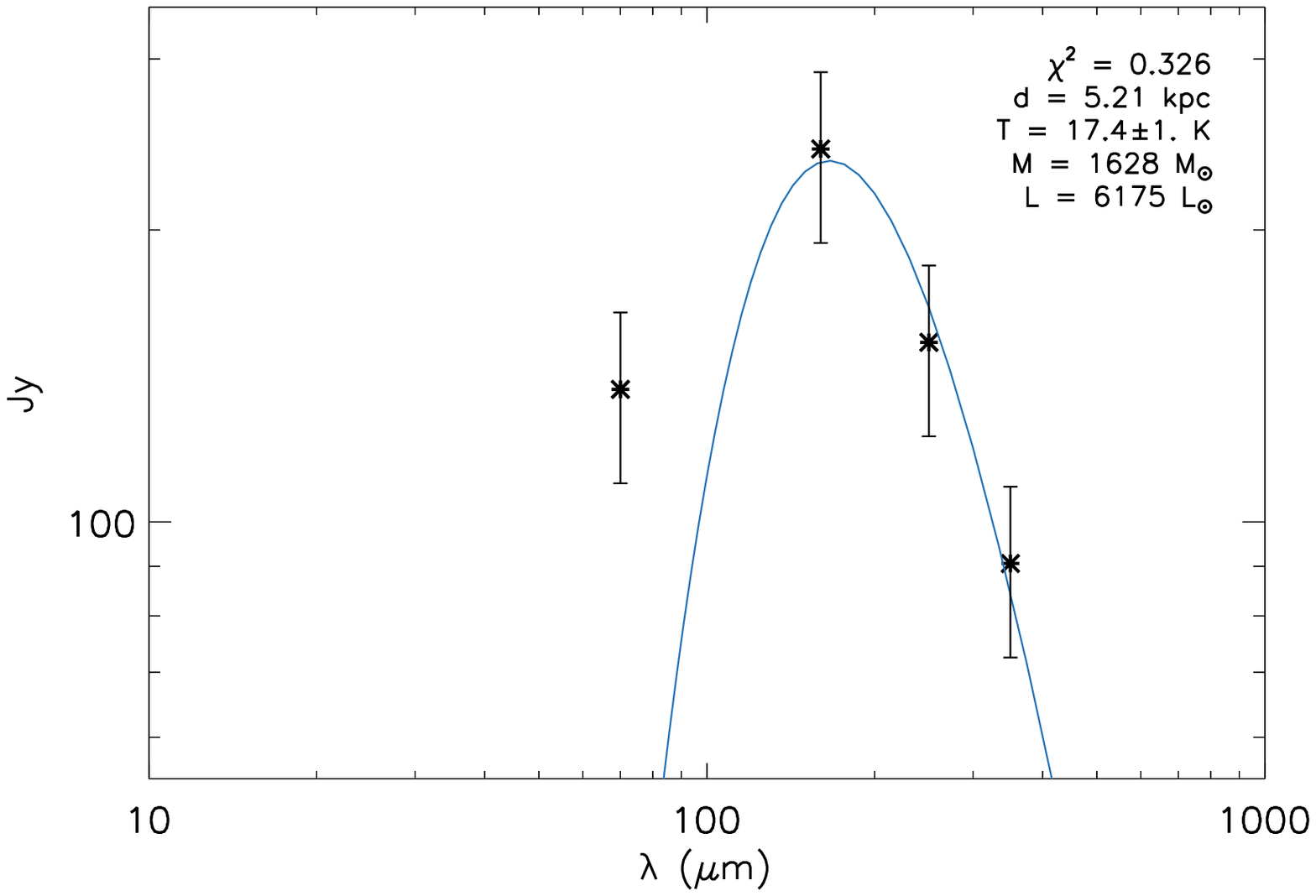}
\includegraphics[width=8cm]{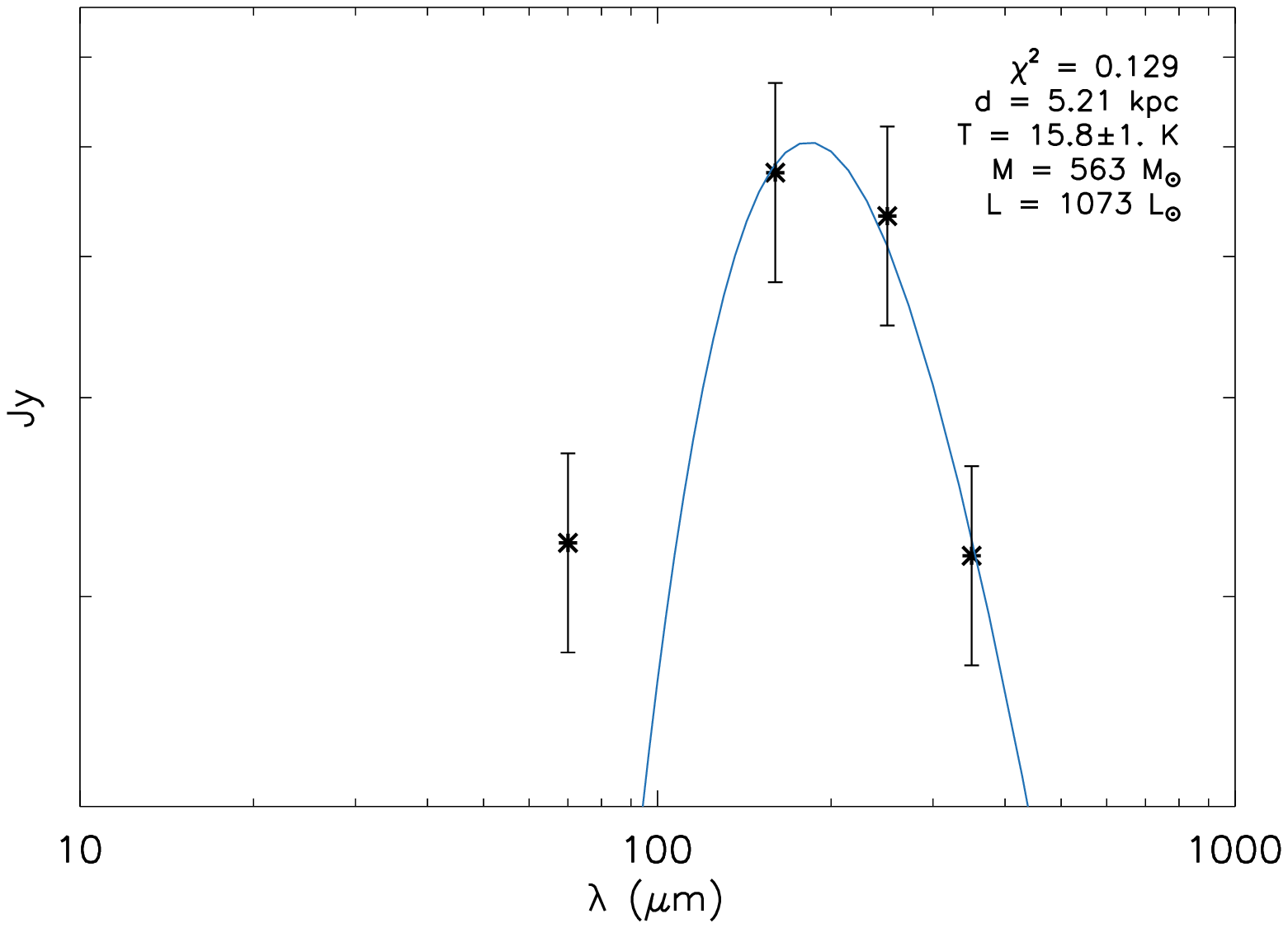}
\includegraphics[width=8cm]{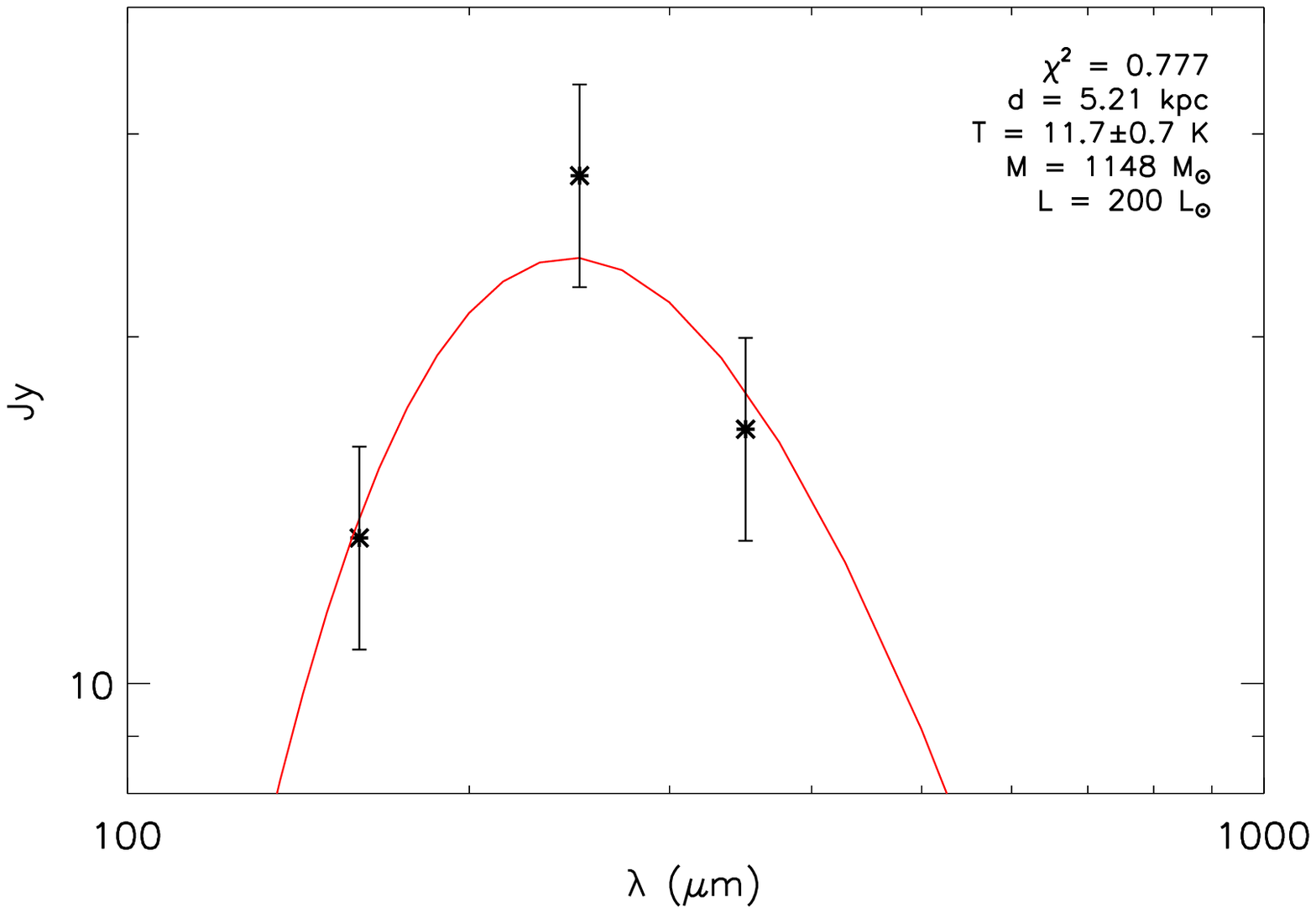}
\caption{Fluxes at 70, 160, 250 and 350 \mum\ and SEDs for the 6 protostellar (blue) clumps and the starless (bottom, red) clumps in HGL23.271-0.263. The spectral index is fixed to $\beta=2.0$. The free parameters of the fit are the temperature and the mass, while the luminosity is obtained integrating the emission in the range $70\leq\lambda\leq500$ \mum. The $\chi^{2}$ value of the fit is also shown.}
\label{fig:SEDs}
\end{figure*}

Figure \ref{fig:SEDs} shows the SEDs for the starless and protostellar clumps observed in HGL23.271-0.263. The plots for the  protostellar clumps also show the emission at 70 \mum, which is in excess of the SED fits in all cases. Each SED plot reports the best-fit parameters, the distance of the source and the value of the $\chi^{2}$. These values are also reported in Table \ref{tab:source_properties}.

\begin{table*}
\begin{center}
\begin{tabular}{c|c|c|c|c|c|c|c|c|c}
\hline
\hline

IRDC & source & type & R & M & L & T & T err & $\chi^{2}$ & distance \\
& & & (pc) & (M$_{\odot}$) & (L$_{\odot}$) & (K) & (K) &  & (kpc)\\ 
\hline

23.271-0.263   &    1   & protostellar &   0.71 &   228   &    926   & 17.7      & 2.1   & 0.595 & 5.21 \\
23.271-0.263   &    2   & protostellar &   0.76 &   1240   &    3110   &  16.0      & 1.6   & 0.055 & 5.21 \\
23.271-0.263   &    3   & protostellar &   0.73 &   152   &    1748   &  22.7      & 3.1   & 1.730 & 5.21 \\
23.271-0.263   &    4   & protostellar &   0.91 &   2537   &   8591   &  17.7      & 1.9   & 0.042 & 5.21 \\
23.271-0.263   &    5   & protostellar &   0.91 &   1628   &    6175   &  17.4      & 1.9   & 0.326 & 5.21 \\
23.271-0.263   &    6   & protostellar &   0.91 &   563   &    1073   &  15.8      & 1.5   & 0.129 & 5.21 \\

\\

23.271-0.263   &    7   & starless &   0.56 &   1148   &    200   &  11.7      & 0.7 & 0.777  & 5.21 \\

\hline
\label{tab:source_properties}
\end{tabular}
\caption{Temperature, mass and luminosity of the 7 clumps identified in HGL23.271-0.263, six protostellar and one starless. Columns 1-2: IRDC name and source number. Column 3: protostellar or starless type. Columns 4--6: radius, mass and luminosity of each source. Columns 7--8: temperature and the associated error estimated with the \texttt{mpfit} routine. Column 9: $\chi^{2}$ value of the greybody fit. Column 10: source distance, equal to the parent IRDC distance.}

\end{center}
\end{table*}

\subsection{Radius, temperature and mass distributions}\label{sec:rad_temp_mass}
We adopt a source size given by the equivalent radius $r_{eq}$, defined as the radius of a circle with the same area, $A=\pi r_{eq}^2$, as the elliptical region evaluated from the 250 \mum\ source fit. Unlike previous work on \Her\ data we do not assume the deconvolved radius to describe the source size, which is a description of the intrinsic shape of the source but it is wavelength-dependent \citep[e.g.][]{Motte10,Giannini12,Elia13}. Instead, $r_{eq}=\sqrt{A / \pi}$ describes the dimension of the region across which the flux is evaluated at all wavelengths.  

\begin{figure}
\centering
\includegraphics[width=9cm]{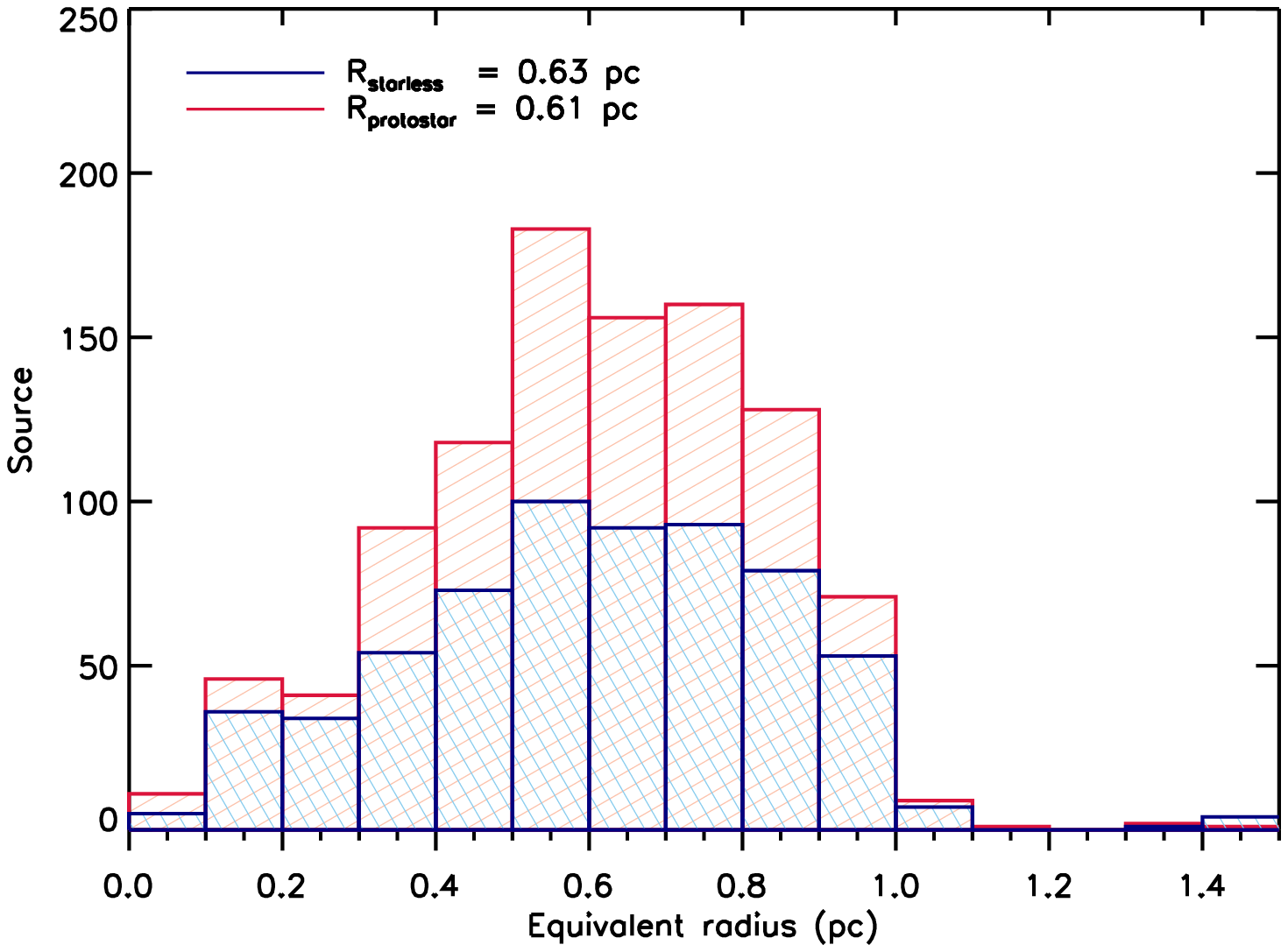}
\caption{Equivalent radius distribution of the clumps identified in the catalogue for both starless (red) and protostellar (blue) clumps. The mean values are 0.61 pc and 0.63 pc for starless and protostellar clumps respectively. The radii have been evaluated as the radii of circles with the same area of the ellipses estimated from the 2d-Gaussian fits at 250 \mum, as explained in detail in the text.} 
\label{fig:radius_distribution}
\end{figure}

The distribution of  $r_{eq}$  for both starless and protostellar clumps is shown in Figure \ref{fig:radius_distribution}. A Kolmogorov-Smirnov (KS) test shows that there is no evidence that the distributions are different. The median values of the two distributions are similar, $\bar{r}_{eq}\simeq0.6$ pc. At the median distance of our sample of IRDCs (4.2 kpc, see Section \ref{sec:source_identification}) the SPIRE 250 \mum\ beam is equivalent to $\simeq0.4$ pc, therefore in most cases we cannot resolve single cores \citep[$r\leq0.1$ pc,][]{Bergin07}. The apparent size threshold around R$\simeq1$\,pc is  because the vast majority of the sources lie below $\simeq5.5$ kpc and we set the maximum aperture radius, which defines the equivalent radius $R_{eq}$ (see Section \ref{sec:rad_temp_mass}), to twice the FWHM$_{250\mu\mathrm{m}}$, 36\arcsec, which corresponds to $R_{eq}\simeq1$ pc at d=5.5 kpc. This choice of restricting the fit to twice the FWHM$_{250\mu\mathrm{m}}$ is made to  isolate compact structures embedded in the clouds and we expect the clumps to be relatively circular, with aspect ratios of less than 2 on average \citep[e.g.][]{Urquhart14}. The few sources with radii $r_{eq}\geq1$ pc are associated with the furthest IRDCs of our catalogue, located around $d\simeq10$ kpc in correspondence of the Perseus arm at $l\simeq50\grad$ (see Section \ref{sec:source_identification}).

\begin{figure}
\centering
\includegraphics[width=9cm]{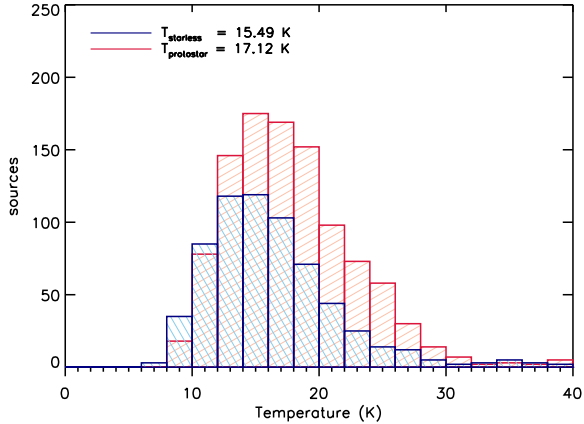}
\caption{Temperature distribution of the 1723 clumps, starless (red) and protostellar (blue). The median values are T=15.49 K and T=17.12 K respectively, suggesting that the cold dust envelopes are slightly warmed by the central core(s) in the protostellar clumps.} 
\label{fig:temperature_distribution}
\end{figure}

The distribution of source temperature for both the starless and protostellar clumps are shown in Figure \ref{fig:temperature_distribution}. The median temperatures of the distributions is T$\simeq$15.5 K and T$\simeq$17.1 K for starless and protostellar clumps respectively. These temperature distributions are in agreement with similar analysis of starless and protostars done with Hi-GAL both in the inner Galaxy \citep[e.g.][]{Veneziani12} and in the outer Galaxy \citep[e.g.][]{Elia13}. The KS test shows that the probability that the two distributions are the same is $<10^{-9}$, although it is clear that the protostellar clumps are only slightly warmer than starless clumps on average. Since the emission in the 160-350 \mum\ spectral range arises from the dust envelope for both starless and protostellar clumps, this result implies that the protostellar envelope is only partially warmed-up by the central core(s), which may be an indication of the youth of the protostellar systems. 

\begin{figure}
\centering
\includegraphics[width=9cm]{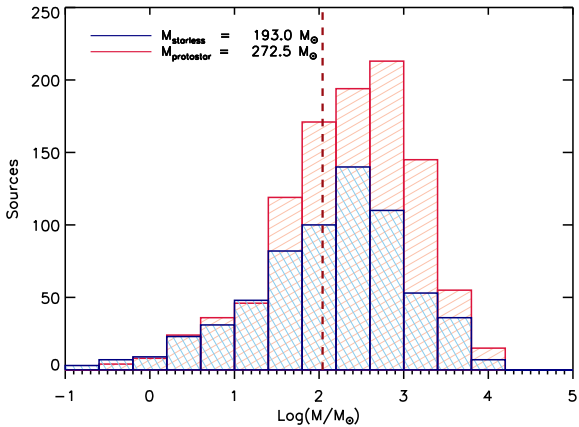}
\caption{Mass distribution of the 1723 clumps, starless (blue) and protostellar (red). The dashed red line is the mass completeness, fixed at M=105 M\sun. The mass range spans several order of magnitudes, from few solar masses up to few $10^{4}$ M\sun. The median values are M=194 M\sun\ and M=269 M\sun\ for starless and protostellar clumps respectively.}
\label{fig:mass_distribution}
\end{figure}

The mass distributions of the sources are shown in Figure \ref{fig:mass_distribution}. The red vertical line defines M$_{com}$=105 M\sun\ which indicates the mass completeness limit as explained in the next Section. This limit lies below the turnover point which is around M$\simeq10^3$ M\sun.
The identified sources span a wide range of masses, from few solar masses up to few $\simeq10^{4}$ M\sun with average clump mass of $\simeq$193 M\sun\ and $\simeq$272 M\sun\ for starless and protostellar clumps respectively. The distributions are similar for both starless and protostellar dumps, likely due to the fact that the envelope of these clumps are not substantially perturbed by the internal gravitational collapse which is occurring in the protostellar cores \citep{Giannini12}. Some of these clumps are potentially the birth site of high-mass objects, as discussed in Sections \ref{sec:mass_radius_diagram} and \ref{sec:mass_surface}. 

Some of the clumps in the sample have also been observed as part of the \Her\ EPOS survey of high-mass star forming IRDCs by \citet{Ragan12} . They found 496 protostellar cores and clumps simultaneously observed at 70, 100 and 160 \mum\ associated with 45 IRDCs across the Galactic Plane. Longer wavelengths were excluded from the analysis in order to preserve the high spatial resolution of the PACS data. In the \citet{Ragan12} catalogue 64 protostellar clumps fall in the longitude range $15\grad\leq l\leq55\grad$ and overlap with our sample. All these 64 sources were detected in our analysis at 70\mum, but only 35 of them have observable counterparts at 160, 250 and 350 \mum\ and are part of the protostellar clump catalogue. In addition 3 sources have been identified as starless, since the 160 \mum\ centroids are located more than $6.7\arcsec$ away from the 70 \mum\ source (Section \ref{sec:source_identification}).

The properties of the two overlapping samples are not straightforward to compare, since the higher spatial resolution of the \citet{Ragan12} sample allows them to model the properties of the inner part of the clumps, and in some cases of the single cores. Indeed in our catalogue the dust temperature are, on average, 5 K colder and the mass are, on average, a factor of $\simeq8$ higher compared to the \citet{Ragan12} catalogue. The mean mass for the 3 clumps we identify as starless is M$\simeq$2900 M\sun\ whereas   \citet{Ragan12} find a mean of M$\simeq740$ M\sun. Similarly for the the 35 protostellar clumps, in our catalogue the mean mass is M$\simeq$1350 M\sun\ whereas  \citet{Ragan12} find M$\simeq$215 M\sun. The inclusion of the 70 \mum\ flux in the SED fitting by \citet{Ragan12} significantly raises the temperature estimation which contributes to these mass differences. The sources have a mean temperature of 14.4 and 18.3 K for starless and 17.5 and 21.7 K for protostellar clumps, in our catalogue and in the \citet{Ragan12} catalogue respectively. 

It is important to note that the mass (and mass surface density) values have significant uncertainties due to the assumed dust model. The opacity and the spectral index of the dust can vary among the different models by a factor of 2 or more \citep{Ormel11}. High-resolution, multi-wavelength observations at sub-mm/mm wavelengths are required to better constraint the dust properties and hence the mass of each clump.

\subsubsection{Mass completeness}\label{sec:flux_completeness}
The mass completeness of the catalogue is determined by the ability to recover faint sources in the presence of highly variable backgrounds, which vary from cloud to cloud by orders of magnitude. Also, we are extracting sources from thousands of relatively small clouds located at a range of distances, $d$, which will have masses $\propto d^{2}$ (see Equation \ref{eq:sed}). Therefore a single completeness limit for the catalogue is poorly representative of the whole sample of clouds. Instead, a mass completeness needs to be evaluated locally for each cloud which accounts for the background confusion and the source distance.

In order to derive the local mass completeness for each cloud we compared the mean mass of the embedded clumps with the corresponding mean \textit{background-equivalent source} mass \bgm\ of each cloud. For each clump we define the background-equivalent source as a point source with a flux equal to the clump background flux integrated in a circular region with a radius equal to the FWHM$_{250 \mu m}$ at all wavelengths (see Section \ref{sec:photometry}). The clump background flux per pixel at each wavelength is defined as the average flux emission per pixel evaluated in a region surrounding each clump, with the clump masked and with a sigma-clipping procedure to avoid significant contamination from companion sources.

The background-equivalent source mass is then evaluated using the same SED fitting procedure described in the previous Section and assuming for the background sources the same distance as their corresponding IRDC. The mean value of all the background-equivalent source masses in each cloud determines the local IRDC \bgm.

\begin{figure}
\centering
\includegraphics[width=9cm]{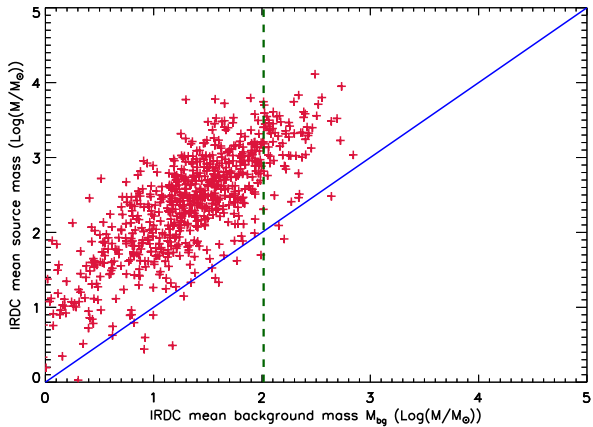}
\caption{Distribution of the mean source mass per IRDC as function of the \bgm. The blue line is the y=x line. The majority of the clumps have M$\geq$ \bgm.}
\label{fig:mass_background_comparison}
\end{figure}

The mean mass of the background-equivalent sources compared with the sources in each IRDC is shown in Figure \ref{fig:mass_background_comparison}. Sources with a mass up to \bgm\ are hidden in the background emission and unidentifiable in that specific cloud. \bgm\ sets a local mass completeness limit which accounts for the local background variation and the IRDC distance.  There are a few tens of clouds with \bgm$\geq{10^2}$ M\sun\ and, at the same time, several clouds for which sources with M$\leq10$ M\sun\ are easily identifiable. On average the mean source mass is few times greater than the corresponding mean \bgm, however there are identified sources with masses close to the corresponding mean \bgm.

To set an estimate of the mass sensitivity limit for the whole catalogue, we identify the mass value M$_{com}$ where 90\% of clouds have \bgm$\leq$M$_{com}$ (the green dashed line in Figure \ref{fig:mass_background_comparison}). In other words, in 90\% of the clouds all sources with masses M$>$M$_{com}$ are detected. This mass value is M$_{com}\simeq$105 M\sun. We define this value as a mass completeness limit for the catalogues. The remaining 10\% of the clouds with \bgm\ above the limit are among the furthest in the catalogue, with a mean distance of 5.4 kpc (in comparison with the average distance of the whole catalogue, 4.2 kpc, see Section \ref{sec:source_identification}).

\begin{figure}
\centering
\includegraphics[width=9cm]{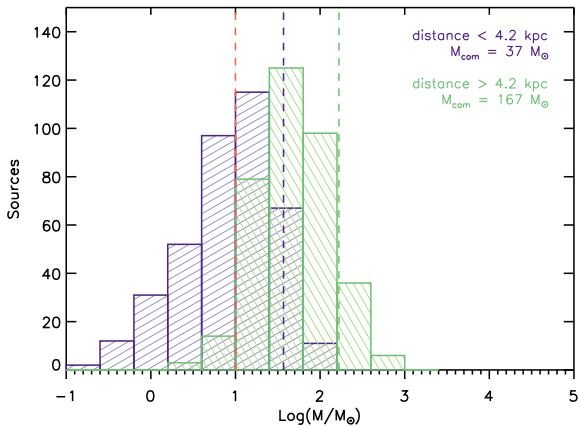}
\caption{\bgm\ distribution for sources located at $d\leq4.2$ kpc (purple histrogram) and at $d\geq4.2$ kpc (green histrogram). M$_{com}$, which defines the mass value above which all sources are identified in 90\% of the clouds, changes significantly with distance with values of M$_{com}$=37 M\sun\ for the distribution of the closest sources and M$_{com}$=167 for the distribution of the more distant sources. The orange-dotted line shows for comparison the mass of a background-equivalent source located at the mean distance of d=4.2 kpc and with an emission at 160 \mum\ arising from residual noise only.}
\label{fig:mass_background_distances}
\end{figure}

We stress that this mass completeness limit is strongly influenced by the IRDC distance and has to be taken ``cum grano salis'', in general a local mass completeness should be used for each cloud. As showed in Figure \ref{fig:mass_background_distances}, the distribution of \bgm\ substantially changes if we divide the source sample in two subsets, one including all the sources located at $d\leq4.2$ kpc and one including all the sources located at $d\geq4.2$ kpc. M$_{com}(d\leq4.2)$=37 M\sun\ and M$_{com}(d\geq4.2)$=167 M\sun\ for the close and the far sample respectively. To further demonstrate that the mass completeness is dominated by the diffuse emission and depends on the source distance, we estimated the mass of a background-equivalent source located at the mean distance of d=4.2 kpc and with an emission at 160 \mum\ arising from residual noise only \citep[0.6 Jy/pixel,][]{Traficante11}. Its mass is $\simeq10$ M\sun\ (the orange-dotted line in Figure \ref{fig:mass_background_distances}). For comparison, the average \bgm\ of all clouds located at $\simeq$4.2 kpc is $\simeq$30 M\sun, while $\simeq10$ M\sun\ is the mean \bgm\ of clouds located at $\simeq2.2$ kpc.

Finally, we note that the selection criteria of our sample of clouds (see Section \ref{sec:dataset}), together with the local mass completeness discussed in this Section, do not produce a complete sample of objects. M$_{com}$ gives a reference value which only defines that we recover sources with masses $\geq$ M$_{com}$ in 90\% of the clouds and it is not related to the underlying intrinsic clump mass spectrum, for which a statistically complete sample would be required. As a consequence, for example, it is not clear whether the turnover point in the source mass distribution, around M$\simeq10^{3}$ M\sun (see Figure \ref{fig:mass_distribution}), is meaningful beyond describing the actual distribution of clumps in the catalogue.
The local mass completeness values for each cloud is available with the source catalogues.

\subsection{Mass vs. radius}\label{sec:mass_radius_diagram}
In the past it has been suggested that IRDCs are favourable places for high-mass star formation \citep[e.g.][]{Carey98}. This suggestion has been corroborated by several observations of massive YSOs associated with IRDCs \citep[e.g.][]{Beuther13,Beltran13,Peretto13}. However, IRDCs are also forming low-to-intermediate stars in regions devoid of high-mass stars \citep[e.g.][]{Sakai13}.

\begin{figure}
\centering
\includegraphics[width=8.5cm]{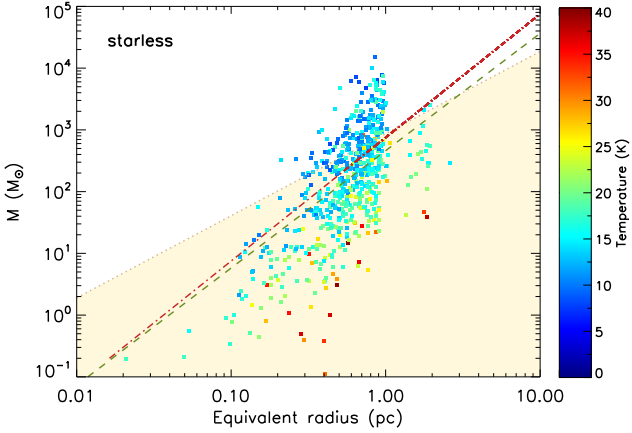}
\caption{The mass of the starless clumps as function of their equivalent radius, colour-coded for the temperature of the dust envelope. The mass increases with increasing equivalent radius. The unshaded area delimits the region of the high-mass star formation in IRDCs as determined by the \citet{Kauffmann10} empiric analysis. Almost one third of the starless clumps (171 out of 667) lie above this threshold. The line-dotted red line is the threshold adapted from \citet{Urquhart14} who propose an empirical lower limit value for high-mass star formation equivalent to a constant mass surface density of  $\Sigma=0.05$ g cm$^{-2}$. The green dashed line is the \citet{Larson81} universal scaling relation between mass and radius of molecular clouds, shown for comparison.}
\label{fig:mass_radius_starless}
\end{figure}

One way to determine if the clumps associated with the IRDCs are ongoing high-mass star formation is to look at the mass-radius distribution under the assumption that, in order to form high-mass stars, a large amount of material needs to be concentrated in to a relatively small volume. Recently an empirical threshold for the formation of mass stars in IRDCs has been proposed. This threshold is, in its original formulation, $m(r)>870$ M\sun\ $(r/\mathrm{pc})^{1.33}$ \citep[][hereafter KP]{Kauffmann10}. This model accounts for single fragments within each star-forming region and it is well suited to be compared with our clumps catalogue. In Figure \ref{fig:mass_radius_starless} we show the mass-radius relationship for our catalogue of starless clumps, colour-coded for their dust envelope temperature. The light brown shaded area delimits the KP region above which massive stars can be formed. For comparison, the figure also shows Larson's third law as it appeared in its original formulation, $m(r)>460$ M\sun\ $(r/\mathrm{pc})^{1.9}$ (green dashed line), which describes the universality of the scaling relation between mass and radius of molecular clouds \citep{Larson81}.

We identify 171 starless clumps above the KP threshold, distributed in 130 IRDCs. Interestingly by this criterion $\simeq$26$\%$ of the starless clumps are likely to form high-mass stars. These are situated in $\simeq$33$\%$ of the IRDCs with identified starless clumps (389, Section \ref{sec:source_identification}). At the same time, the clouds which encompass these high-mass starless clumps are only 4$\%$ of the total clouds in our sample (3493, see Section \ref{sec:dataset}). While the mass completeness limit influences the percentage of high-mass starless clumps, the number of clouds which embed these high-mass clumps is much less affected since it is unlikely that we are missing the most massive objects associated with the clouds. We can conclude that only a small fraction of IRDCs will potentially form high-mass stars, in agreement with the results of \citet[][]{Kauffmann10}.

The dash-dotted red line in Figure \ref{fig:mass_radius_starless} defines the empirical threshold for high-mass star formation proposed by \citet{Urquhart14}. This threshold is based on the analysis of high-mass star forming clumps identified in the ATLASGAL survey and corresponds to clumps with a constant mass surface density of  $\Sigma=0.05$ g cm$^{-2}$. This threshold however sets less stringent constraints than the KP threshold for clumps with an equivalent radius of R$\leq1$ pc, which are the majority of the clumps in our survey. We discuss in the next Section the mass surface density thresholds for high-mass star formation compared with our sample.

\begin{figure}
\centering
\includegraphics[width=8.5cm]{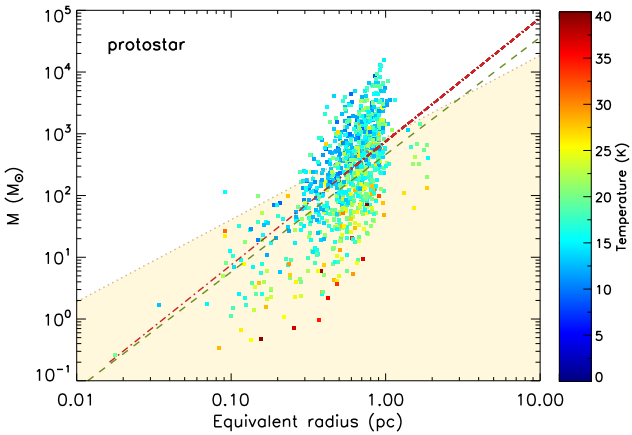}
\caption{The mass of the protostellar clumps as function of their equivalent radius, colour-coded for the temperature of the dust envelope. There are $\simeq$43$\%$ of the protostellar clumps (456 out of 1056) above the \citet{Kauffmann10} threshold. }
\label{fig:mass_radius_protostar}
\end{figure}

The mass-radius diagram for protostellar clumps is shown in Figure \ref{fig:mass_radius_protostar}. A direct comparison of the protostellar clump sample with the KP threshold is less meaningful however, since the mass estimation of protostellar clumps can be affected by mass-loss in jets or outflows, or mass segregated in the central core(s) as well as overestimating the mass as a result of using a single temperature model and incompletely accounting for the heating by the central protostars. However, we find 456 protostellar clumps ($\simeq$43$\%$ of the total) associated with 291 IRDCs ($\simeq$50$\%$ of the IRDCs with embedded protostars, and $\simeq$8$\%$ of the total) above the KP threshold and so are capable of forming high mass stars. 

The low percentage of IRDCs possibly forming high-mass stars does not exclude however that the majority of the high-mass star formation activity is associated with IRDCs. This issue can be addressed, for example, by comparing the star formation activity within the IRDCs and in a large sample of massive molecular clouds not associated with IRDCs, or tracing the star formation rate inside and outside IRDCs. Both these analyses are the main topic of a forthcoming work.



The temperature distribution in the diagrams show that the colder clumps are the bigger and the more massive. For starless clumps this may be an indication of their evolutionary phase, with the smaller clumps having already started central accretion and likely being just at the verge of a protostar core formation, or maybe with a very young protostar already formed but still obscured at 70 \mum. For protostellar clumps this is likely a consequence of the adopted model, since the single clump-averaged temperature is dominated by the extended cold dust envelope emission with less contribution from the warmer central regions as the clump size increases.

\subsection{Mass and Mass surface density}\label{sec:mass_surface}
The environment of massive star formation can be also investigated through the mass surface density, $\Sigma$, versus mass diagram. The work of  \citet{Krumholz08} suggested $\Sigma=1$ g cm$^{-2}$ as a mass surface density threshold required to prevent fragmentation into low-mass cores as a result of radiative feedback, thus allowing high-mass star formation. However, this threshold has a large scatter and the calculation does not include the contribution of magnetic fields, which can prevent fragmentation without imposing a minimum mass surface density \citep[e.g.][]{Butler12}. Several high-mass clumps and cores are indeed observed with $0.1\leq\Sigma\leq0.5$ g cm$^{-2}$ \citep{Butler12,Tan13}. One of the most massive cores observed in the inner Galaxy, associated with the IRDC SDC335, with a mass of $\simeq500$ M\sun\ in $\simeq0.05$ pc \citep{Peretto13} reaches $\Sigma\simeq60$ g cm$^{-2}$. However the mass surface density of the embedding clump, with a mass of $\simeq2600$ M\sun\ in $\simeq1.2$ pc \citep{Peretto13}, is $\Sigma\simeq0.5$ g cm $^{-2}$. The dark cloud, with a mass of $\simeq5500$ M\sun\ within $\simeq2.4$ pc has a mean surface density of $\Sigma\simeq0.25$ g cm $^{-2}$. On average, high-mass star forming regions associated with IRDCs are usually observed within $0.1\leq\Sigma\leq1$ g cm$^{-2}$ \citep{Tan14}. As discussed in the previous Section, a less stringent empirical threshold for high-mass star formation in clumps has been suggested by \citet{Urquhart14} based on ATLASGAL clump studies and corresponds to $\Sigma\geq0.05$ g cm$^{-2}$.

\begin{figure*}
\centering
\includegraphics[width=16cm]{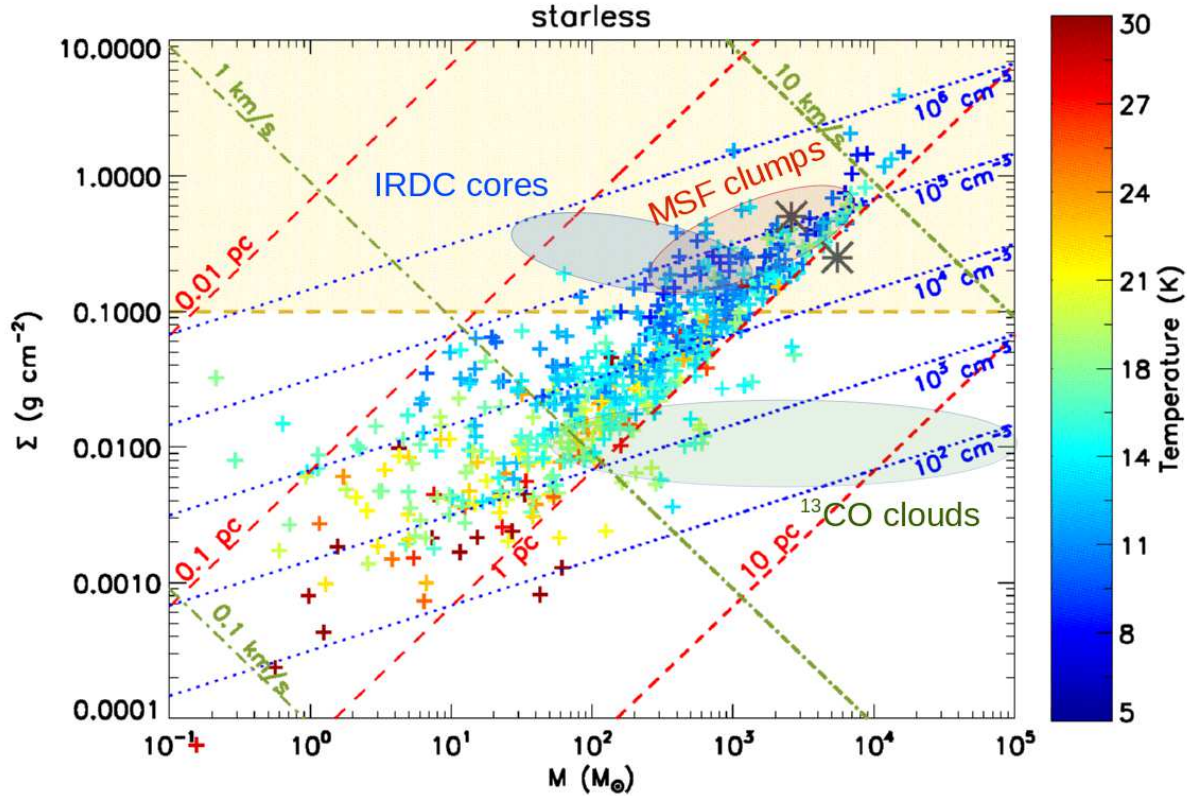}
\caption{Mass surface density as function of mass for starless clumps, colour coded for the temperature of the dust envelope. The light brown shaded area defines the region of sources with surface density $\Sigma\geq0.1$ g cm$^{-2}$. The  lines correspond to constant values of various parameters: \textit{red lines}: constant radius; \textit{blue lines}: constant H number density, analogous of constant free-fall time; \textit{green lines}: constant escape velocity. Also shown are the regions corresponding to IRDC cores observed by \citet[][blue shaded area]{Butler12}, the massive star formation clumps (and cores) observed by \citet[][red shaded area]{Mueller02} and the $^{13}$CO molecular clouds from the catalogue of \citet[][green shaded area]{Roman-Duval10}. The dark grey asterisks show the location of the SDC335 central clump and cloud \citep{Peretto13}.}
\label{fig:surface_density_starless}
\end{figure*}

\begin{figure*}
\centering
\includegraphics[width=16cm]{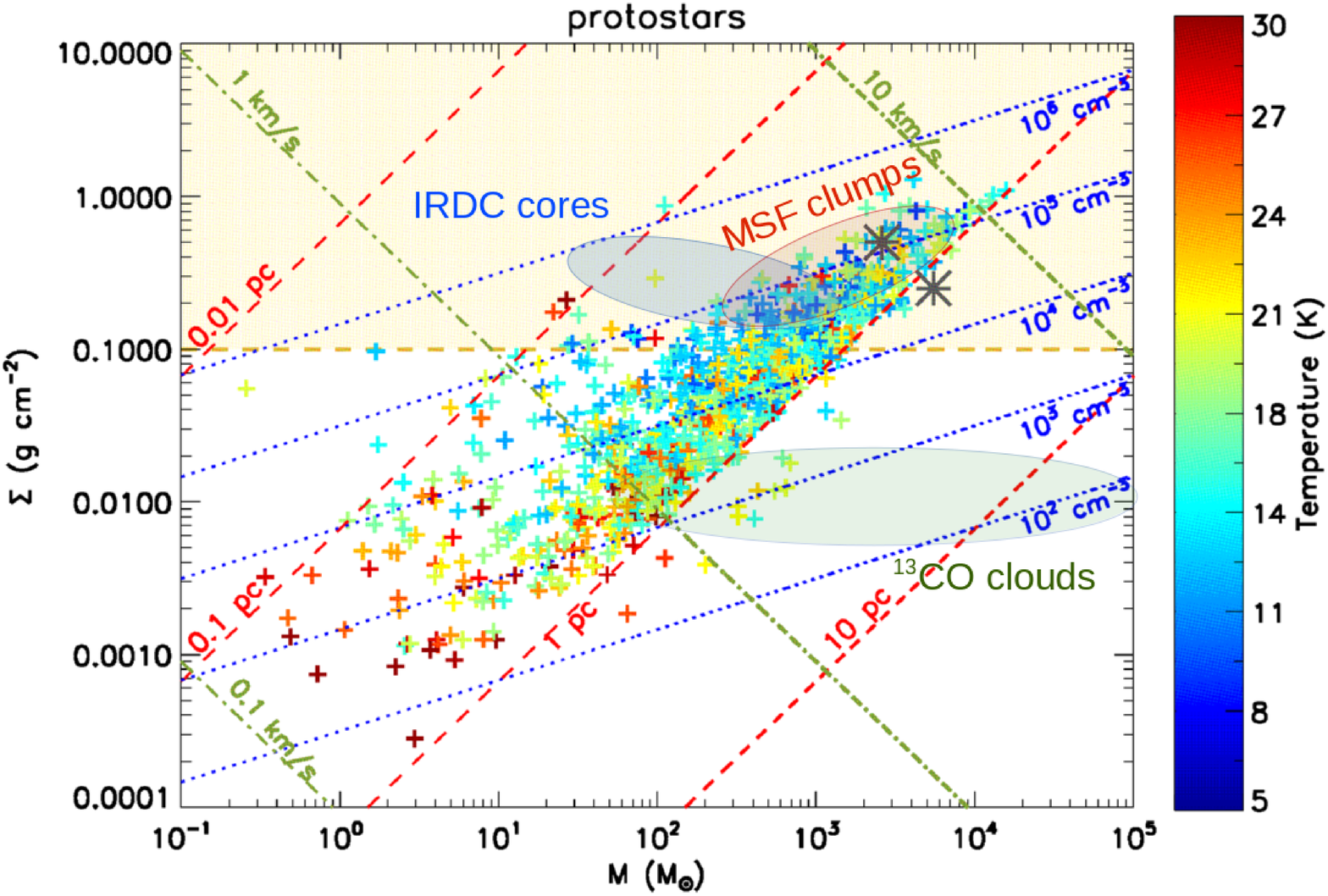}
\caption{Mass surface density as function of mass for protostellar clumps, colour coded for the temperature of the dust envelope. The shaded areas and lines are as shown in Figure~\ref{fig:surface_density_starless}.}
\label{fig:surface_density_protostar}
\end{figure*}

In Figures \ref{fig:surface_density_starless} and \ref{fig:surface_density_protostar} we plot the mass surface density-mass diagrams for the starless and protostellar clumps respectively, colour-coded for the temperature of the dust envelope. The light-brown shaded area delimits the $\Sigma\geq0.1$ g cm$^{-2}$ region. There are 144 starless and 308 protostellar clumps, respectively 22\% and 29\% of the two samples, within this region where massive star formation is possible. Adopting the \citet{Urquhart14} threshold limit increases these numbers to  241 starless and 542 protostellar clumps ($\simeq$36\% and $\simeq$51\% of the respective samples). Following \citet{Tan14}, we also show straight lines corresponding to constant radius, constant escape speed and constant hydrogen number density. The apparent radius threshold at R$\simeq1$ pc is due to our choice of the maximum aperture size of clumps and the distance distributions of clouds, as discussed in Section \ref{sec:rad_temp_mass}.

The escape speed is $v_{esc}=(10/\alpha_{vir})^{1/2}\sigma$, with $\alpha_{vir}$ being the virial parameter and $\sigma$ the velocity dispersion of the clump. Assuming a virial parameter $\alpha_{vir}=1$ for gravitationally bounds clumps and an average velocity dispersion of $\sigma=1$ km/s \citep[e.g.][]{Henshaw13,Kauffmann13} the escape speed is $v_{esc}\simeq3$ km/s. Therefore, for a typical high-mass star forming clump with mass of M$\simeq 1000$ M\sun\ \citep{Tackenberg12, Urquhart14}, the clump is bound if its mass surface density reaches $\Sigma\simeq0.1$ g cm$^{-2}$. Low values of viral parameters have been observed in region of high-mass star formation \citep[e.g.][]{Tan13,Li13}. Assuming $\alpha_{vir}=0.5$ the corresponding escape speed becomes  $v_{esc}\simeq4.5$ km/s and a M$\simeq 1000$ M\sun\ clump will be bound only if reaches $\Sigma\simeq0.3$ g cm$^{-2}$ or, equivalently, if its mass is confined in a radius R$\simeq$ 0.45 pc. We find 38 starless and 76 protostellar clumps with $\Sigma\ge0.3$ g cm$^{-2}$, $\simeq6$\% and $\simeq7$\% of the two samples respectively. We find only 4 starless and 4 protostellar clumps with $\Sigma\ge1$ g cm$^{-2}$. 

The hydrogen number density $n_{H}$ is analogous to constant free-fall time, following the relation $t_{ff}=[3\pi/(32G\rho)]^{1/2}$,  where $\rho$ is the volume density of the clump. From the diagram we note that, setting the high-mass star formation threshold at $\Sigma\geq0.3$ g cm$^{-2}$, the minimum hydrogen number density of high-mass star-forming clumps is $n_{H}=10^{5}$ cm$^{-3}$. This value corresponds to a free-fall time of $1.4\times10^{5}$ year. Assuming as a lower limit for the mass surface density $\Sigma\geq0.1$ g cm$^{-2}$ to form high-mass stars, from the diagram we deduce a minimum hydrogen number density of $n_{H}\simeq1.5\times10^{4}$ cm$^{-3}$, or a free-fall time of $3.6\times10^{5}$ year. With the surface density proposed by \citep{Urquhart14}, from our clumps distribution we deduced a minimum hydrogen number density of $n_{H}\simeq8\times10^{3}$ cm$^{-3}$, or a free-fall time of $5.0\times10^{5}$ year.


For comparison we highlighted the regions in the plots occupied by cores/clumps associated with a sample of 10 IRDCs selected by \citet{Butler12}, the 51 massive star forming clumps identified by \citet{Mueller02} and the $^{13}$CO molecular clouds identified in \citet{Roman-Duval10}. The SDC335 central clump and cloud values are also showed in the plot with dark grey asterisks.



\subsection{Luminosity}\label{sec:luminosity}
\begin{figure}
\centering
\includegraphics[width=9cm]{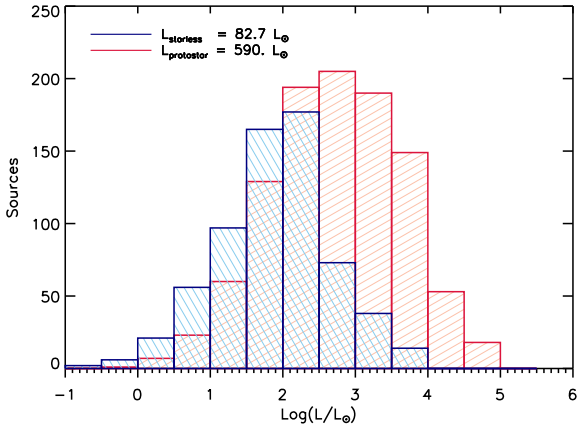}
\caption{FIR luminosity distributions for the 1723 clumps, evaluated in the range $70\leq\lambda\leq500$ \mum\ as detailed in the text. The median values are L=83 L\sun\ and L=590 L\sun\ for starless and protostellar clumps respectively. This difference is a direct consequence of the internal luminosity of the core(s) embedded in the protostellar clumps which are partly visible in the 70 \mum\ emission.}
\label{fig:luminosity_distribution}
\end{figure}

Figure \ref{fig:luminosity_distribution} shows the FIR luminosity distributions of the clumps. 
The median values of the luminosities are L=83 \Lsun\ and L=590 \Lsun\  for starless and protostellar clumps respectively. This difference in the FIR luminosity is principally due to the emission arising from the warm core(s) in the protostars. In the starless clumps the luminosity of the dust envelope is the only contribution to the bolometric luminosity. On the other hand, the warm core(s) in the protostellar clumps contributes substantially to the integrated FIR luminosity. Depending on the evolutionary state of a protostar there can also be significant emission at 24 \mum\ or shorter wavelengths. The estimated FIR luminosity is therefore a lower limit of the protostars bolometric luminosity \citep[e.g.][]{Giannini12}. 

 We identified 71 protostellar clumps ($\simeq$7$\%$ of the total) with L$\geq10^{4}$ \Lsun. There are 52 starless clumps with L$\geq10^{3}$ \Lsun, relatively high for starless clumps. The majority of them (30 out of 59) have a bolometric luminosity vs. envelope mass (L$_{bol}$ / M$_{env}$)$\leq1$, indicative of objects without warm cores. These are among the most massive starless clumps identified in this catalogue. The remaining 22 starless clumps with luminosity L$\geq10^{3}$ \Lsun\ and L$_{bol}$ / M$_{env}>1$ are either very young protostellar clumps or not confirmed starless, as from visual inspections some of them show them to be are associated with particularly faint 70\,\mum\ sources not identified with our selection threshold (see Section \ref{sec:source_identification}) or associated with bright diffuse 70\,\mum\ emission.


\subsection{Mass-Luminosity}
A potentially useful tool to infer the evolutionary properties of the young clumps is the L$_{bol}$ -- M$_{env}$ diagram. This diagram has been successfully used in the past to describe the evolutionary phases of low-mass objects \citep{Saraceno96}, and subsequently extended to predict the evolutionary paths of high-mass objects \citep{Molinari08}. A source evolves following specific tracks, depending on its initial mass and luminosity, and the model predicts the mass and luminosity of the object when it reaches the zero age main sequence (ZAMS). For high-mass objects the paths follow the two-phases model of \citet{McKee03}. According to this model, when a cloud starts its gravitational collapse the mass slightly decreases due the accretion and molecular outflows, while the luminosity of the object increases significantly, sustained by the collapse. The source moves along almost vertical paths in the diagram. At the end of this phase the object is surrounded by an HII region, the luminosity remains constant while the mass is expelled through radiation and molecular outflows. The object then follows an almost horizontal path, which will ends when it becomes visible in the optical and almost all the envelope has been expelled. Accordingly to the evolutionary phase in the diagram, the high-mass objects are classified in analogy with the low-mass regime classification, from Class 0 to Class II prior to the ZAMS phase. This classification can be misleading though, and some caveats of this model have to be kept in mind, in particular when applied to our sample: 1) the transition from Class 0 to Class I and Class II sources in the low-mass regime is not sharp. Sources classified as Class 0 based on their NIR--MIR fluxes can instead be classified as Class I in the FIR \citep{Hennemann10,Elia13}; 2) the evolutionary tracks have been initially modelled for single cores, not for clumps and 3) in high-mass regions the observed clumps can contain multiple cores in different stages of evolution and the clump luminosity is likely dominated by the most evolved core(s).  
Nevertheless, the L$_{bol}$ -- M$_{env}$ diagram has been used in the past to discuss the evolutionary tracks of YSOs with \Her\ data in a wide range of different conditions, both for low-mass \citep[e.g.][]{Bontemps10,Hennemann10} and for intermediate-to-high mass regime \citep[e.g.][]{Elia10,Veneziani12}. 
 
\begin{figure}
\centering
\includegraphics[width=9cm]{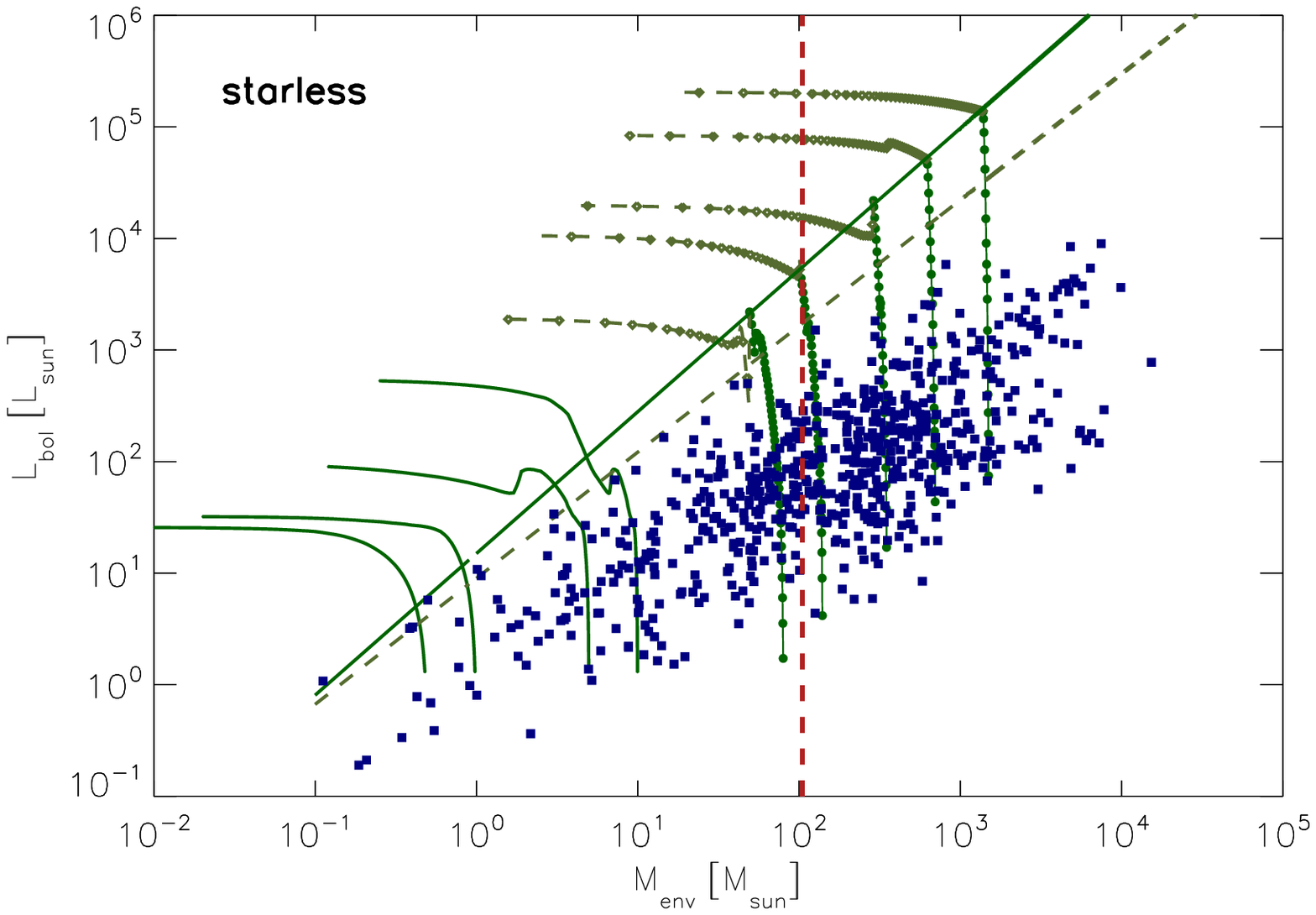}
\caption{L$_{bol}$ -- M$_{env}$ diagram for starless clumps superimposed with the model of \citet[][]{Saraceno96} and \citet[][green lines]{Molinari08}. The solid and dashed lines are the best log-log fit for Class I and Class 0 sources respectively, extrapolated in the high-mass regime by \citet{Molinari08}. The red vertical line is our mass completeness limit, M=105 M\sun. The sources are sparse in the diagram but the majority are below the Class 0 fit, showing that they are in a very early stage of evolution.}
\label{fig:mass_lum_starless}
\end{figure}

\begin{figure}
\centering
\includegraphics[width=9cm]{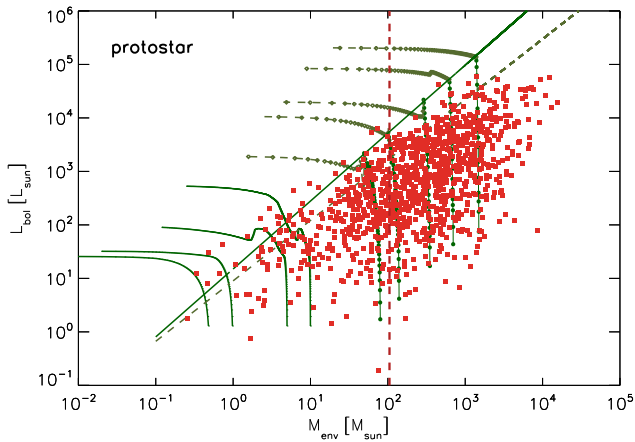}
\caption{The same diagram of Figure \ref{fig:mass_lum_starless} but for protostellar clumps.}
\label{fig:mass_lum_protostar}
\end{figure}

The L$_{bol}$ -- M$_{env}$ diagram for the starless clumps is shown in Figure \ref{fig:mass_lum_starless}. The distribution is superimposed to the models of \citet{Saraceno96} and \citet{Molinari08}, which shows the tracks for stars with final masses of 6.5, 8, 13.5, 18 and 35 M\sun\ from left to right respectively. The best log-log fit for Class I and Class 0 high-mass objects \citep{Molinari08} are also shown. The majority of the sources lie below the best log-log fit of Class 0 objects. Also, the distribution lies on average below the distribution of young high-mass clumps identified in \citet{Urquhart14}, which includes clumps at different stages of evolution but does not include starless ones. We find a mean L$_{bol}$ / M$_{env}\simeq1.1$, compared with a mean L$_{bol}$ -- M$_{env}\simeq10$ of \citet{Urquhart14}, consistent with the majority of these starless clumps being in a very early stage of their evolution. Figure \ref{fig:mass_lum_protostar} shows the L$_{bol}$ -- M$_{env}$  diagram for the protostellar clumps. The distribution overlaps that of the starless clumps, proving little discrimination between the relative evolutionary stages of the sources. However the mean L$_{bol}$ -- M$_{env}\simeq6.6$ is an indication of a more evolved phase and including measurements at shorter wavelengths will increase the luminosity of many sources (Section \ref{sec:mass_luminosity}) and raise the distribution in the diagram. In addition, higher angular resolution measurements which can resolve the actual envelope mass (as opposed to the clump mass) associated with each protostar, will move the points to the left. 
Without these additional data, the high degree of overlap in the distributions limits the value of these plots for probing the evolutionary status of the sources.

\subsection{Clumps lifetime}
Comparing the number of starless to protostellar clumps provides an
estimate of the lifetime of the starless phase. This is of particular
interest for those regions most likely to form high mass stars, the
most massive clumps with high surface densities. For clumps with a
mass M$>500$\,\Msol\ and a surface density $\Sigma>0.1$\,g cm$^{-2}$ the ratio
of the number of starless clumps to protostellar clumps is 0.46, which
increases only slightly to 0.48 when considering clumps with a surface
density $\Sigma>0.3$\,g cm$^{-2}$. This suggests a lifetime for the starless
phase of about half that of the protostellar phase.  Adopting an
estimate of few$\times 10^5$\,yr for the lifetime of the embedded
phase of young massive stars \citep[e.g.][]{Davies11}, the starless phase duration would
be $\sim10^5$ yr, in agreement with the lifetime estimation of starless clumps identified in the ATLASGAL survey \citep{Tackenberg12}.  Since some starless clumps may contain currently
unidentified protostars as suggested by the high luminosity to mass
ratio for some of the apparently starless clumps (Section \ref{sec:luminosity}), this
lifetime represents an upper limit.

\section{Conclusion}\label{conclusion}
We have produced a catalogue of starless and protostellar clumps identified in a sample of 3493 IRDCs extracted from the \citet{Peretto09} catalogue in the region of the Galactic Plane delimited by $15\grad\leq l\leq55\grad$, $\vert b\vert\leq 1\grad$. Using the \Her\ observations acquired as part of the Hi-GAL survey, we first identified the  FIR counterparts of the IRDCs in the Hi-GAL data at 70, 160, 250 and 350 \mum. The compact sources were extracted from the Hi-GAL IRDCs counterparts using \Hyp, a new algorithm which combines the advantages of the aperture photometry with a minimal use of source modelling to take into account the high variable background and the source crowding \citep{Traficante14_Hyp}. The sources were initially identified at 160 \mum\ and we found 5393 160 \mum\ sources distributed in 1640 IRDCs ($\simeq$47\% of the initial sample). All the sources with a counterpart at 250 and 350 \mum, but no counterpart(s) at 70 \mum\ are identified as \textit{starless}. If at least one 70 \mum\ counterpart is present, they are identified as \textit{protostellar} clumps. We identified 1056 protostellar in 586 IRDCs ($\simeq$17\%) and 667 starless clumps in 389 IRDCs ($\simeq$11\%).  There are 211 IRDCs which contain both starless and protostellar clumps.

Assuming for each source the distance of its parent cloud we evaluated the equivalent radius, temperature, mass and luminosity of the sources dust envelope, modelling the Hi-GAL fluxes in the range 160-350 \mum\ with a single-temperature greybody model and a fixed spectral index of $\beta=2.0$. The adopted model is a good approximation to describe the cold envelopes with only  small temperature gradients across the clumps, but it does not account for the emission arising from the central, warmer regions of protostars. As a consequence it underestimates the average dust temperature (and therefore overestimates the mass) of the protosteallr clumps. The mean equivalent radius is $\simeq0.6$ pc for both the starless and protostellar objects, representative of clumps. This is probably driven by our limited capacity of resolving smaller clumps of cores. Indeed the mean distance of the clumps, 4.2 kpc, does not allows us to resolve objects smaller than $\simeq0.4$ pc on average. The distributions of the source temperatures are different, as demonstrated by a KS test, however the mean temperature values differ for less than 2 K, T=15.5 K and T=17.1 K for starless and protostellar clumps respectively. The mass distributions are also similar, with median values of 193 M\sun\ and 272 M\sun\ respectively, and extending up to $\simeq10^4$ and $\simeq2\cdot10^4$ M\sun\ for the more massive starless and protostellar clumps respectively. The similarities in temperature and mass indicate that the dust envelope of the protostellar clumps has been not yet warmed-up by the inner core(s) and it is relatively unaffected by the gravitational collapse occurring in the core(s). Both these conclusions indicates that the population of protostellar clumps observed in this sample of IRDCs is very young. 

We demonstrated that, due to the heterogeneity of the cloud structures, the mass completeness has to be assumed locally for each cloud. We have however set a reference mass completeness of M=105 M\sun\ for the whole sample, which corresponds to the mass above which clumps can be detected in 90\% of the clouds. 

The mass vs. radius distribution of the starless clump is used to identify the clumps which may lead to high-mass stars, assuming the \citet{Kauffmann10} empiric threshold for high-mass star formation. We show that almost one third of the starless clumps (171 out of 667), distributed in 130 IRDCs (only 4\% of the whole sample), are above the high-mass stars threshold. In other words only a small fraction of the clouds are likely to form massive stars.

The environment of massive star formation has been further investigated through the mass vs. surface density diagram. Assuming that high-mass star formation occurs in regions with $\Sigma\geq0.1$ g cm$^{-2}$ \citep{Butler12,Tan14} we identify 144 starless and 308 protostellar clumps that will likely form high-mass stars. From the mass surface density diagram of the starless clumps we identify the maximum free-fall time of our sample of t$\simeq4.4\times10^5$ year. Fixing the mass surface density threshold to $\Sigma\geq0.05$ g cm$^{-2}$, as proposed by \citet{Urquhart14}, this time rises to t$\simeq5.0\times10^5$ year. The numbers of massive (M$\geq500$ M\sun), high surface density starless
clumps and those already containing embedded sources indicate an upper
limit lifetime for the starless massive clump phase of $\sim10^5$
years, similar to the free-fall time.



The FIR luminosity distributions differs with mean values of L=83 L\sun\ and L=590 L\sun\ for starless and protostellar clumps respectively, principally because the protostellar clumps luminosity is highly influenced by the luminosity of the embedded core(s). A total of 71 protostellar clumps, $\simeq7\%$ of the sample, have FIR luminosity $\geq10^{4}$ L\sun. The FIR luminosity is a good approximation for the bolometric luminosity only for starless clumps, and we compared the starless L$_{bol}$-- M$_{env}$ diagram with the evolutionary models for low- and high-mass objects developed by \citet{Saraceno96} and \citet{Molinari08}. The starless clumps distribution is sparse but most of the sources are below the line which describes the high-mass counterparts of low-mass Class 0 regime. Although a clear analogy with the low-mass classes classification is not straightforward in the high-mass regime, the mean L$_{bol}$ / M$_{env}\simeq1.1$ of the distribution shows that the starless clumps are in the very early stages of their evolution.

These catalogues provide a significant sample to identify interesting sources to study in detail the mechanisms involved in the early stages of  star formation. Future work will extend the catalogue to the whole sample of IRDCs in the Galactic Plane and will constraint the dust properties of these clumps, identifying counterparts at shorter wavelengths, as well as in the high-resolution sub-mm/mm surveys of the Galactic Plane such as ATLASGAL and the BGPS.

The catalogues are freely available at the webpage \texttt{http://www.irdarkclouds.org}.

\section*{acknowledgements}
AT and GAF acknowledge the support from the STFC consolidated grant
ST/L000768/1.


\bibliographystyle{mn2e}
\bibliography{bibliography.bib}

\label{lastpage}

\end{document}